\documentclass[10pt]{article}
\usepackage{graphics,epsfig}

\font\smallroman=cmr9

\begin{document}

\bigskip

\bigskip

\bigskip

\bigskip

\title{\textbf{{\sc Interpreting the Quantum Wave Function \\in Terms of `Interacting Faculties'}}}
\author{\sc Christian de Ronde}
\date{}
\maketitle \centerline {Center Leo Apostel (CLEA), and Foundations
of the Exact Sciences (FUND),} \centerline {Faculty of Science,
Brussels Free University} \centerline {Krijgskundestraat 33, 1160
Brussels, Belgium.} \centerline {cderonde@vub.ac.be}

\bigskip

\bigskip

\bigskip

\bigskip

\bigskip

\bigskip

\begin{abstract}
\noindent In this article we discuss the problem of finding an
interpretation of quantum mechanics which provides an objective
account of physical reality. In the first place we discuss the
problem of interpretation and analyze the importance of such
an objective account in physics. In this context we
present the problems which arise when interpreting the quantum
wave function within the orthodox formulation of quantum
mechanics. In connection to this critic, we expose the concept of
`entity' as an {\it epistemological obstruction}.

In the second part of this paper we discuss the relation between
actuality and potentiality in classical and quantum physics, and
continue to present the concept of \emph{ontological potentiality}
which is distinguished from the generic Aristotelian notion of
potentiality in terms of `becoming actual'. In this paper our main
aim is to provide an objective interpretation of quantum mechanics
which allows us to discuss the meaning of physical reality
according to the theory. For this specific propose we present the
concept of \emph{faculty} in place of the concept of `entity'.
Within our theory of faculties, we continue to discuss and
interpret two paradigmatic experiments of quantum
mechanics such as the double-slit and Schrodinger's cat.
\end{abstract}

\noindent

\newtheorem{theo}{Theorem}[section]

\newtheorem{definition}[theo]{Definition}

\newtheorem{lem}[theo]{Lemma}

\newtheorem{prop}[theo]{Proposition}

\newtheorem{coro}[theo]{Corollary}

\newtheorem{exam}[theo]{Example}

\newtheorem{rema}[theo]{Remark}{\hspace*{4mm}}

\newtheorem{example}[theo]{Example}

\newcommand{\proof}{\noindent {\em Proof:\/}{\hspace*{4mm}}}

\newcommand{\qed}{\hfill$\Box$}

\newcommand{\ninv}{\mathord{\sim}} 

\newpage

\tableofcontents

\bigskip

\bigskip

\begin{flushright}
\emph{Perhaps in the future one shall have to rethink what it\\
 really means to get at the end of an experiment a number,\\
that one has to compare with the number predicted by the theory.}\\
\vspace{0.5cm}
Diederik Aerts.
\end{flushright}

\section*{Introduction}

This article deals with the problem of interpretation in quantum
mechanics. This problem has haunted physics since the very
beginning of the last century when Max Plank made a small change
in the equation which governed the emission of radiation of a
black body. The shift from a \emph{continuous} conception of
energy to a \emph{discrete} one cleared a path which classicality
had overlooked. Still today we are trying to understand the
meaning of the theory of quanta, its possibilities and
impossibilities. Its possibilities regard not only technical
developments but also philosophical viewpoints, the
impossibilities expose the limits of our classical conception of
the world. In this paper we intend to propose an interpretation of
quantum mechanics which can provide an objective account of
physical reality. The price we are willing to pay is the
re-consideration of what is to be considered `physically real'
according to quantum mechanics.

\section{The Problem of Interpretation}

In the year 1900 Max Planck expressed the idea that energy is not
continuous, but rather, comes in discrete packages called {\it quanta}.
At this very moment the departure from our classical conception of
the world begun. The conception which had been worked out since
Plato and Aristotle till Newton and Kant, was now for the first
time, seriously threatened. The critics came from different
scopes: philosophy, music, poetry and even physics tackled the
metaphysical conception of the world at the end of the 19th
century. The nature of the problems raised by the new quantum
theory revealed a critical confrontation with the classical
worldview. When Schr\"odinger wrote his famous `cat article' in
1935 it was already clear that the return to the departed land was
not possible anymore.

{\smallroman
\begin{quotation}
``[...] if I wish to ascribe to the model at each moment a
definite (merely not exactly known to me) state, or (which is the
same) to {\small {\it all}} determining parts definite (merely not
exactly known to me) numerical values, then there is no
supposition as to these numerical values {\small{\it to be
imagined}} that would not conflict with some portion of quantum
theoretical assertions." E. Schr\"odinger (\cite{Schrodinger35},
p.156)
\end{quotation}}

As physicists we are interested in giving a picture of the world,
a story that puts together all which we have experienced through
creation and discovery. Although we have learned a lot since this
critical period the `classical conception of the world' has
remained our single view, that through which we understand and
describe our existence.\footnote{As Constantin Piron has pointed
out the revolutions of relativity and quantum have not yet taken
place \cite{Piron}.} This view has certain presuppositions which
confront the basic structure of the quantum formalism, making it
very difficult ---maybe even impossible--- to paste it altogether with
our previous classical ideas. However, after endless discussions
regarding the meaning of the quantum, in the year 2000, exactly
one century after the beginning of the voyage, Christopher Fuchs
and Asher Peres finally settled the question: {\it Quantum Theory
Needs no `Interpretation'.}

{\smallroman
\begin{quotation}
``[...] quantum theory does not describe physical reality. What it
does is provide an algorithm for computing probabilities for the
macroscopic events (``detector clicks") that are the consequences
of experimental interventions. This strict definition of the scope
of quantum theory is the only interpretation ever needed, whether
by experimenters or theorists." C. Fuchs and A. Peres
(\cite{FuchsPeres00}, p.1)\end{quotation}}

\noindent Of course this emphasis in prediction in detriment to
description can be severely questioned. The main objection against
this instrumentalistic point of view is that the success of a
theory can not be explained, that is to say, we do not know how
and why quantum physics is in general able to carry out
predictions (and in particular with such a fantastic accuracy).
Undoubtedly a ``hard" instrumentalist may simply refuse to look
for such an explanation, since it is in fact the mere
effectiveness of a theory that which justifies it, so that he may
not be interested in advancing towards a justification of that
effectiveness. If one takes such a position there is nothing left
to say. Just like the Oracle of Delphos provided always the right
answer to the ancient Greeks, quantum mechanics provides us with
the correct probability distribution for every experiment we can
think of. So there is nothing else we need; that is all we can
ask from a physical theory and there is no need to supplement it
with an \emph{interpretation}. This end to the discussion was
something which Werner Heisenberg and Wolfgang Pauli could have
not foreseen, even though they had called the attention over such
an elusive answer on the question of interpretation many years
before:

{\smallroman
\begin{quotation}
``Wolfgang Pauli hizo esta observaci\'on: ``El silencio no se debi\'o
a que tu explicación fuese mala. Pertenece, en efecto a la
profesi\'on de fe del positivismo el que deben aceptarse los hechos
reales sin reparo alguno. Si mal no recuerdo, Wittgenstein afirma
aproximadamente lo siguiente: `el mundo es todo aquello que
sucede', `el mundo es el conjunto de los hechos, no de las cosas.'
Cuando se admite este postulado como punto de partida, es forzoso
admitir sin vacilaci\'on una teor\'ia que representa tales hechos. Los
positivistas saben que la mec\'anica cu\'antica describe con exactitud
los fen\'omenos at\'omicos; por consiguiente, no tienen motivo alguno
para oponerse a ella. Todo lo que los f\'isicos a\~{n}adimos despu\'es,
como, por ejemplo, complementariedad, interferencia de
probabilidades, relaciones de indeterminaci\'on, diferencia entre
sujeto y objeto, etc., todo esto les parece a los positivistas un
lirismo carente de claridad, un retorno al pensamiento
precient\'ifico, pura charlataner\'ia. De todos modos, no hay que
tomarlo en serio y, en el mejor de los casos, resulta inofensivo.
Quiz\'a tal concepci\'on constituya en s\'i misma un sistema l\'ogico
cerrado. Pero yo no entiendo entonces qu\'e queremos decir cuando
decimos `comprender la naturaleza'." W. Heisenberg (\cite{Heisenberg72}, p.255)\end{quotation}}

I believe that physics is not merely about algorithms, it is not
only a predictive machinery, its importance cannot be reduced to
that of being `a source of technical developments'. Physics is the
gentle communion between {\it mathematical expressions}, {\it description} and {\it experience}.
Its basic presupposition is that Nature {\it exists}, and we, as
physicists, want to discuss what Nature {\it is}. Physics is a
particular way of studying the Being, this is why it involves {\it
epsitemology} on the one side, and {\it ontology} on the
other.\footnote{For a more detailed discussion in relation to this
position see \cite{deRondeCDI, deRondeCDII}.} Ontology is for us a
picture, an abstract conceptualization, a story of a world which
we seek to understand. As we have discussed in detail in
\cite{deRondeUQMCD, deRondeRel}, in quantum mechanics, it is the
development of the ontology which remains a central issue:

{\smallroman
\begin{quotation}
``When the layman says ``reality" he usually thinks that he is
speaking about something which is self-evidently known; while to
me it appears to be specifically the most important and extremely
difficult task of our time to work on the elaboration of a new
idea of reality." W. Pauli (\cite{Laurikainen98}, p.193)
\end{quotation}}

This is the problem of providing meaning to our theories, the
problem of understanding, which is at the same time the problem of
interpretation.

\subsection{An Objective Account of Physical Reality}

Objectivity is one of the fundamental cornerstones of science. This is the very democratic principle which governs it: everybody can check by himself, with the proper means, a certain aspect of a theory. This attitude draws the most important distinction with other approaches towards the problem of reality. In particular, religious or mystic views of reality depart from this democratic conception; for example, it is not necessarily true that everybody can be enlightened or see a miracle. To have a vision is a personal experience, maybe not even communicable to someone else. Contrary to science, religious and mystic approaches to the world in which we live are based on personal experience.

We believe that science is committed to objectivity. By this we mean that, firstly, their propositions must be robust under {\it intersubjective agreement}.\footnote{D'Espagnat has called this mode of reference: {\it weakly objective} statements. See for example  \cite{D'Espagnat06} p. 98.} Everyone should be able to check a certain character of a theory, the empirical elements of a certain theory should not be viewed only by a `chosen one'. For example, the phenomena discussed in the theory of gravitation, by which a body is attracted to the earth, should be (in principle) robust under the checking of anyone who wishes to perform an experiment which tests the theory. Most of miracle-type experiences do not follow this presumption, for example the existence of ghosts is not an inter-subjective experience, generally speaking there are only gifted people or mediums which get in contact with ghosts; i.e. it is not necessary true that everyone can see a ghost. In second place, in order for a theory to be objective, we need a set of elements, represented in the mathematical formulation of the theory, which provide access to {\it empirical recognition} and are robust under certain {\it consistency constrains} provided by the ontology to which the theory is committed. The consistency constrains deal with the mode of existence of the elements under investigation. Together, all this structure provides meaning to physical reality and tells us what the theory is about, this is over which we must agree and disagree as physicists.

However, it is important to remark that intersubjective agreement is a necessary but {\it not sufficient condition} to define physical objectivity. Physics is not just a consistent discourse about phenomena, physics does not only talk about directly observable phenomena, and this is why, it has been directly engaged with the history of metaphysics. We tend to accord with Einstein who expressed very clearly the guiding line of physics:

{\smallroman
\begin{quotation}
``[...] it is the purpose of theoretical physics to achieve understanding of physical reality which exists independently of the observer, and for which the distinction between `direct observable' and `not directly observable' has no ontological significance; this aim furnishes the physicist at least part of the motivation for his work; but the only decisive factor for the question whether or not to accept a particular physical theory is its empirical success." A. Einstein (Quoted from \cite{Dieks88}, p.175)
\end{quotation}}

It seems quite common nowadays to state a kind of reluctant view on the relation between physical reality and the world. There are many positions which seem to propose a new trend of physics which does not have anything to do with ``physical reality". The importance of technical developments is closely related to this new conception which seems to forget the problems and presuppositions of doing physics. The question raises: if we are not talking about ``the world", about ``physical reality", then what is physics about?

Niels Bohr's ideas have played a central role in the development of physics in the 20th century, placing the discipline within the main philosophical line of discussion of the period, i.e., the problem of language and its relation to ontology and epistemology.\footnote{The linguistic turn refers to a major development in Western philosophy during the 20th century, the most important characteristic of which is the focusing of philosophy on language as constructing reality \cite{Scavino}. The relation between the linguistic turn and Bohr's development is discussed in \cite{deRondePhD}.} Maybe the most clear statement about this point is the famous quotation by Aage Petersen. According to his long time assistant Bohr once declared when asked whether the quantum world could be considered as somehow mirroring an underlying quantum reality:

{\smallroman
\begin{quotation}
``There is no quantum world. There is only an abstract quantum
physical description. It is wrong to think that the task of
physics is to find out how nature is. Physics concerns what we can
say about Nature" N. Bohr quoted by A. Petersen
(\cite{Petersen63}, p.8) \end{quotation}}

Language appears here not only as the regulative notion of access, but as the limit and final goal of physics itself. Or as Bohr used to say: ``We are suspended in language in such a way that we cannot say what is up and what is down. The word `reality' is also a word, a word which we must learn to use correctly.''\footnote{Quoted from: Philosophy of Science, 37, 1934, p. 157.} In Bohr's writings there is no reference to ontology, nor to physical reality. His analysis remains in this point extremely superficial turning quantum mechanics into a theory of knowledge about classical phenomena. Objectivity is viewed by Bohr as intersubjective agreement, only in terms of a linguistic discourse: ``The description of atomic phenomena has in these respects a perfectly objective character, in the sense that no explicit reference is made to any individual observer and that therefore... no ambiguity is involved in the communication of observation."\footnote{N. Bohr quoted from \cite{D'Espagnat06} p. 98.} We will return to this point later in section 6.1.\\

In early Greek thought the study of Nature was approached through
\emph{contemplation}, one could gain true knowledge of a Cosmos
which was meant to exist. Physics was born from this idea, which
was later engaged with more specific conceptions. Maybe the most
important presupposition in physical thought, which can be traced
to the reading of Parmenides by Plato and Aristotle, is that the
Cosmos is constituted by \emph{entities}. This idea was formally
expressed by Aristotle in his logic, which was nothing but the
building blocks of our classical conception of
Nature.\footnote{See for discussion \cite{VerelstCoecke}.} Much
later, it was Isaac Newton himself who was able to translate into
a closed mathematical formalism the ontological presuppositions
present in Aristotelian logic. Classical physics is thus, nothing
but the study of preexistent entities, of things which exist. This
is the ideal of classical thought which we must clearly recognize
in order to understand not only its power but also its
limitations.

It is through the conception of the Being, as an entity, that
science has approached the world since Plato and Aristotle. Let it
be a particle, a wave or a field, an entity is a primitive concept
of the theory which preexists to our analysis of its existence. In
the case of quantum mechanics the problem is raised because of its
apparent subjective character with respect to the entities under
study; i.e. the mode of Being of elementary particles. In quantum theory the epistemic foundation based in the
ideas of classical physics appears to be at stake, this becomes
clear from Heisenberg's 1927 paper (\cite{Heisenberg27}, p.187) in
which he writes: ``I believe that one can formulate the emergence
of the classical `path' of a particle pregnantly as follows: {\it
the `path' comes into being only because we observe it.}" In
quantum mechanics there is an incompatible structure which
precludes the possibility of thinking of the preexistence of
properties, if one is willing, at the same time, to keep the
categorical structure of entities.\footnote{I have discussed this
point in more detail in \cite{deRondeCDII} and in
\cite{deRondeRel}.}

In quantum mechanics, in order to define the entity under study, we need to choose a cut and objectivity is regained only once the measuring apparatus is chosen. As expressed by Pauli: ``[...] there remains still in the new kind of theory an \emph{objective reality}, insamuch as these theories deny any possibility for the observer to influence the results of a measurement, \emph{once the experimental arrangement is chosen.}"\footnote{Quoted from \cite{Pauli94}, p.33, emphasis added.} Even though, reproducibility of the result is then obtained in the form of objective probability laws for series of repeated measurements, the problem remains untouched. The problem deals explicitly with the dependence of the choice of the context in relation to the determination of its existence. Einstein was worried exactly about this point, about the impossibility of quantum mechanics to refer to an objective account of physical reality. As Wolfgang Pauli recalled in a letter to Bohr dated February 15, 1955:

{\smallroman
\begin{quotation}
``[...] `Like the moon has a definite position' Einstein said to
me last winter, `whether or not we look at the moon, the same must
also hold for the atomic objects, as there is no sharp distinction
possible between these and macroscopic objects. Observation cannot
{\small{\it create}} an element of reality like position, there
must be something contained in the complete description of
physical reality which corresponds to the {\small{\it
possibility}} of observing a position, already before the
observation has been actually made.' [...]" W. Pauli
(\cite{Laurikainen88}, p.60)
\end{quotation}}

Niels Bohr followed a different path, he wanted to regain
objectivity by watching quantum theory from a distance, standing
on the well known heights of the classical scheme. But, if the
objective character of quantum theory ought to be secured by the
use of the classical language, we certainly get into a vicious
circle. We take it that the objective character of a theory should
be secured by the theory itself, or, in case it is an appendix of
a different theory, one should clearly understand their
relationship. As it stands, the position of Bohr forces us into a
very unclear relation between the classical world and the quantum
formalism, which does not seem to have a place in the classical
conception of the world, but nevertheless, talks about it.

This is the problem which is raised in quantum theory: we have the
need to give some picture of what is going on, and explain at the
same time, why is it that we can predict with such an accuracy so
many phenomena. The natural attitude towards the problem, specially after the second half of the 20th century, has been
to force the interpretation into our known classical scheme. Such
proposals, like for example Bohmian mechanics, Many Worlds and
GRW, presuppose the classical conception of the world (or a
similar variant of it), and then try to fit the formalism of the
theory into this pre-conceived idea of reality. On the contrary, we
consider that the term `reality' should not be taken as a
primitive in the task of science, rather, it should be conceived
as a goal conceptualization. Physical reality should not be a
pre-established concept nor a prejudice in observing and relating
empirical data, but rather a goal concept which should be
transformed and developed. We should not expect reality to be...
as we would like it to be; but we must constantly revise the
conceptual framework with which such a description is expressed.
Trying to understand reality (in physical terms) presupposes that we do not know much
about it. It is this very humble attitude which must guide
science. Our strategy will be to seek for the objective part of
the structure in the quantum formalism; forgetting for a moment
the ontological commitments. We have to remind ourselves we are
only scientists wondering about the world, a very mysterious world
indeed.

\subsection{Interpreting Quantum Mechanics}

Quantum mechanics was developed in the first three decades of the
20th century by, mainly, the German speaking community of
physicists. The discussions did not stop between them, not even
between those who seemed to share a common view ---as for example
between Niels Bohr, Wolfgang Pauli and Werner Heisenberg--- until
the late '50. These discussions were certainly of most importance
for the development of the theory. The unfortunate term
``Copenhagen interpretation" invented by Heisenberg in 1958 and
used today by the average physicist as a set of weird rules of
which one should not ask too much, has helped to silence these
discussions \cite{Howard}. A close reading of these authors is
enough to discover there is no single interpretation; their ideas
and philosophical grounds remain different and even incompatible
in substantial points. Instead of a ``Copenhagen interpretation" I
believe it would be more appropriate to refer to the ``Copenhagen
discussions". These discussions were concentrated, firstly, in the
problem of how to make sense of the formalism in relation to
language, through neo-Kantian philosophy in the case of Bohr, and secondly, to the problem of reality, through
Plato's philosophy in the case of Heisenberg \cite{vW85}, and
liked to the philosophy of Schopenahuer, in the case of
Pauli \cite{Laurikainen88}.

Because we understand quantum mechanics as providing a radical
departure of our classical conception of the world, our attitude
regarding its interpretation will remain radical as well. We take
quantum mechanics very seriously, we believe that its formalism
must be understood in terms of a language capable of expressing
what this theory has to tell us about reality. Paul Dirac
(\cite{Dirac47}, p.10) stated that: ``[...] the main object of
physical science is not the provision of physical pictures, but is
the formulation of laws governing phenomena and the application of
these laws to the discovery of new phenomena. If a picture exists,
so much the better; but whether a picture exists or not is a
matter of only secondary importance." On the contrary, we think
that the importance of pictures does not have a secondary place in
physical theories. It is through such pictures that physicists are able to provide
new ideas and to create new experiences. These pictures are for
the physicists guiding lines without which we would sink, no hope
or possibility to approach a new land would remain without these
lighthouses. Exactly this kind of guide is lacking in quantum
mechanics. There is a state of affairs in physics which might be
regarded as analogous to the situation of special relativity at
the end of the 19th century. The formalism of the theory is
finished, the experiments which confirm it have been done, but
still, our unwillingness to give up our classical classical
conception of the world has lead us into all sort of `ether-type
solutions' such as: ether-fields, ether-worlds, ether-jumps, etc.

In order to provide a suitable framework for a proper analysis and
development of the interpretation of quantum mechanics we have
presented the complementary descriptions approach
\cite{deRondeCDI, deRondeCDII, deRondeUQMCD}. A first point of
departure of this framework is a development of the idea of
complementarity, which must refer not only to complementarity
between mutually incompatible phenomena, but also to
complementarity between incommensurable  descriptions. A {\it
description} is a general framework in which concepts relate,
concepts and relations which determine the precondition to access
a certain expression of reality. A description involves creation,
it is a condition of possibility for experiencing and
understanding, it expresses a limit under which it is possible to
provide meaning. But a limit is at the same time an expression of
something which lies outside, an externity which is not
recognized, which must be forgotten. This is why our development
deals specifically with an exposition of the limitations of a
definite description. The complementarity of descriptions is an
expression regarding the impossibility of \emph{reduction} and the
acceptance of the \emph{limitations} implied by the frameworks
themselves. In this line of investigation we have presented the
thesis that classical physics and quantum physics are
complementary descriptions which involve mutually incompatible
concepts and pictures.

Let us be clear about this point, a physical theory is a whole which discusses, through a mathematical formulation and a conceptual scheme, a certain expression of the Being. The Being is {\it expressed} by the theory in terms of its conceptual structure and its mathematical formulation. Against naive realism, physics is not about {\it strongly objective} statements ---which refer directly to some attributes of the things under study--- as D'Espagant calls them \cite{D'Espagnat06}. Physics regards the particular region just in between mathematical expressions, conceptual description and experience.\footnote{The ontology that we'd like to put forward is closely related to Spinoza, through which the one is expressed in different modes, by the many representations. It should be stressed here that in the Spinozist ontology there is no pluralism involved, avoiding the main problem of relativism into which Putnam's internalism is dragged into by the linguistic turn \cite{deRondeSpinoza}.}\\

The core of the problem regarding the interpretational issues of quantum theory are an explicit expression of a more fundamental one, the problem of existence. What does it mean that something exists? Classical thought had forgotten this question which must be reviewed in quantum physics. We are ready now to make a detour and continue beyond the limits imposed by our classical conception of the world.

\section{Interpreting the Quantum Wave Function}

The most basic problem in quantum mechanics remains to interpret
the quantum wave function. What is $\Psi$? What does it represent?
We understand this question as independent of another one, which
is taken as equivalent by many: What does a {\it superposition}
mean? In earlier works we have called the attention over a clear
distinction which must be taken into account for a proper
discussion of these subjects. This distinction is between what we
have called earlier `perspective' and `context' \cite{deRonde03}.

In the orthodox formulation of quantum mechanics the wave function is expressed by an abstract mathematical form. Its representation can be expressed through the choice of a determined basis $B$. The non-represented wave function is a {\it perspective} which expresses pure {\it potentia}, the potentiality of an action which makes possible the choice of a definite context. A perspective expresses the power for a definite representation to take place, it deals with the choice between mutually \emph{incompatible} contexts. The perspective cannot be written, it shows itself through the different representations, each of
which is a part but not the whole. The importance of defining this
notion is related to the structure of the quantum formalism which,
contrary to classical mechanics, is essentially holistic and thus,
intrinsically contextual; i.e., it does not allow for the
simultaneous existence of mutually incompatible contexts. The {\it
context} is a {\it definite representation} of the perspective, it
depends and configures in relation to the concepts which are used
in the description. The different possible contexts can not
be thought as encompassing a whole of which they are but a part
(see \cite{deRondeCDII}, section 1.2).

It is only at the level of the context that one can speak of {\it
properties}. Different set of properties arise in each
representation, which relate and are configured by the
\emph{logical principles} which govern the description. In the
case of classical mechanics properties relate via the principles
of classical (Aristotelian) logic; i.e. the {\it principle of
existence}, the {\it principle of identity} and the {\it principle
of non-contradiction} (see section 4). The reductionistic
character of the structure arises from the choice of these
ontological principles, which at the same time, allows us to speak
of ``something" which exists. It is only because one presupposes
this structural configuration that one is allowed to speak about
{\it entities}. In quantum mechanics the properties arising in
each context do not follow classical relationships but are
determined by a different logic. Heisenberg's {\it principle of
indetermination}, Bohr's {\it principle of complementarity} and
the {\it superposition principle} provide a structural
relationship between quantum-properties which cannot be subsumed
into classical thought.\footnote{For a detailed analysis and
discussion of the principles of indetermination, complementarity
and superposition as those which determine the fundamental logical
structure of quantum theory see \cite{Lahti80}.} When speaking of
properties, one must recognize the discourse in which they are
embedded. In many discussions regarding the interpretation of
quantum mechanics one talks about quantum and classical properties
just like ``properties" without a proper mention to its \emph{mode
of being}, this lack of clarification produces lots of
pseudo-problems and misunderstandings which have been discussed
earlier (see \cite{deRondeCDII}, section 2).

The perspective has not been acknowledged in quantum mechanics due to the metaphysical presuppositions which involve the characterization of a quantum state as a vector in Hilbert space. In orthodox quantum mechanics it is assumed that the {\it vector} ``exists'', independently of the basis in which it is `placed'. That which we need to discuss is the relation between a mathematical expression and physical existence. For the mathematician the definition of a vector has a clear expression and presents no problem whatsoever. However, when discussing the physical interpretation of the vector as representing a ``state of a system'' things get more foggy. In the mathematical structure of quantum mechanics the basis plays an {\it active} role, it constitutes the existence of the set of properties which, at a later stage, determines that which will be studied. ``That which will be studied", and can be best characterized by a {\it superposition}, can not be subsumed into the classical categories of an `entity'. Physically, it is assumed that the $\Psi$ contains all the different representations as existents, this means that all the representations can be captured together in terms of an identity, a unity. The $\Psi$ is something which thus, should be able to give an account of the totality of the different representations as showing parts of a {\it sameness} \cite{GdeR07}. But as we know, specially through the Kochen Specker (KS) theorem \cite{KS}, the ``same" vector cannot support the existence of its different representations simultaneously, precluding in this way the possibility of thinking of $\Psi$ in terms of something which refers to an entity. As we will show in this paper, it is exactly this idea which cannot be maintained in quantum mechanics.

The distinction between perspective and context was introduced in \cite{deRonde03} in order to distinguish between the different modes of existence of the properties in the modal interpretation. Following van Fraassen's distinction between {\it dynamical state} and {\it value state}, we have distinguished between {\it holistic context} and {\it reductionistic context}. This distinction regards the path from a superposition to an ensemble, from an improper mixture to a proper mixture. Notice that one might talk in a reductionistic context {\it as if} one would have an entity, provided that we forget the procedure of successive cuts by which we arrived from the holistic context with improper mixtures. In such case we might talk {\it as if} these mixtures were proper, and thus recover the logical principles which allow us to talk about entities. This interpretational jump has no justification whatsoever, but remains necessary for the later interpretation in terms of probability. (see \cite{deRondeCDII}, section 1.4)

A brief outline of what we tried to explain until now can be
provided in the following scheme:\\

\begin{tabular}{|c|c|c|c|c|}

\hline
& \tiny{\textbf{PERSPECTIVE}} & \textbf{\tiny{HOLISTIC}} & \tiny{\textbf{REDUCTIONISTIC}} & \tiny{\textbf{MEASUREMENT}} \\
&  & \textbf{\tiny{CONTEXT}} & \tiny{\textbf{CONTEXT}} & \tiny{\textbf{RESULT}} \\
\hline
\tiny{\textsl{MATHEMATICAL EXPRESSION}} & $\Psi$ & $|\psi_{B}\rangle$ & $|\psi_{B}\rangle$ & $\alpha_{k}$, $|\alpha_{k}\rangle$ \\
\hline
\tiny{\textsl{CONCEPTUAL EXPRESSION}} & -- & \footnotesize{{\it superposition}} & \footnotesize{{\it ensemble}} & \footnotesize{{\it single term}}\\
\hline
\tiny{\textsl{PROPERTY}} & -- & \footnotesize{{\it holisic/non-Boolean}} & \footnotesize{{\it reductionistic/Boolean}} & \footnotesize{{\it actual}} \\
\hline
\end{tabular}

\subsection{The Idea of `Entity' as a First Approximation}

Since Aristotle presented his logic, the idea of {\it entity}
became the guiding line of physical thought. The idea of entity is
based in the ontological principles of logic presented by
Aristotle as a solution to the problem of movement. It is through these
principles that an entity is capable of unifying, of totalizing in
terms of ``sameness". It is the idea of entity which generated the
development of physics since Aristotle, this is why we might say
today that {\it the history of classical physics is the history of
physical entities.}

As noted by Carl Friedrich von Weizs\"ascker \cite{vW74}, Martin
Heidegger discussed in {\it Sein und Zeit} the problem of
confronting the classical conception of the world, a world which
Plato and Aristotle had created for us. It is since Plato that
occidental thought has confused the `Being' with the `entity'. The
confusion resides in asking about `entities' instead of asking
about the `Being itself', entitazing the Being. In other words, in
asking always about the Being in terms of an entity, not allowing
for the Being to exist in a different form than that provided by
the idea of entity. This mistake was repeated once and again not
only in philosophy but also in physics; in this sense Alfred North Whitehead was correct
to point out that the history of western philosophy has remained
nothing but footnotes to Plato.

That which plays the r\^ole of an entity in a physical theory must
have a counterpart in the mathematical formulation. In the case of
quantum mechanics we have a very good mathematical formulation but
no consistent interpretation. So we can ask ourselves, as a first
beat, if the quantum wave function $\Psi$ can be considered as
referring to an entity. If $\Psi$ is the one to play our staring
r\^ole in the theory, there are certain conditions which it must
fulfill. First of all, it must be robust under different
transformations, these transformations should allow us to state
consistently that that which we transform remains the `same'. The
most obvious examples are given by the Galilean transformations in
classical mechanics, and the Lorentz transformations in special
relativity. In quantum mechanics this might seem to work at first
sight, because $\Psi$ is considered to be nothing but a vector in
Hilbert space, and thus, by definition, an invariant (something
which does not change under rotations). But, as we shall see,
things get very tricky...

In special relativity theory a context is given by a definite
inertial frame of reference. However, there is no need of defining
the perspective because the invariance principle, given by the
Lorentz transformations, allows us to think all these different
contexts as existing in {\it actuality}, as frameworks for events which pertain
to physical reality. There is a way by which one can relate all
the events which actually exist in the same picture (even though
their relation is different of course form that of classical mechanics). In
quantum mechanics, on the other hand, a context is given by a
definite experimental set up which is mathematically represented
by a complete set of commuting observables (C.S.C.O.),
equivalently defined by a quantum wave function in a definite
representation/basis. But because of \emph{Bohr's principle of
complementarity} there is an intrinsic, ontological incompatibility
between different representations. The KS theorem, to which we will return later, does not allow to think
a property, which is seen from different contexts, as existing in
actuality. Thus, the different contexts can not be thought in
terms of possible views of one and the same ``something", namely, the $\Psi$. In
classical mechanics and special relativity theory, this problem
does not arise because one can relate contexts through the
Galilean and Lorentz transformations. One may say that in these
theories one can reduce all the different views to a \emph{single
context},\footnote{Even if we do not know the context we can think
in terms of possible contexts, in terms of ignorance.} and this is
why the idea of perspective becomes superfluous. The formal
structure of classical mechanics and relativity theory is
reductionistic, and thus, part and whole can be univocally
related.

We have provided the distinction between the perspective, in which we have the non represented mathematical form $\Psi$, and the context,  in which a particular representation is expressed through the choice of a basis $B$ and we can write the wave function as $|\psi_{B}\rangle$. We believe that this distinction is very important in order to clarify the problem we are discussing. The `weird structure' of quantum mechanics avoids that we can talk of the different representations $|\psi_{B}\rangle$, $|\psi_{B'}\rangle$, $|\psi_{B''}\rangle$, ... as views of ``the same" $\Psi$. This is due to the fact, that the choice of the basis plays an {\it active} role in the definition of that which exists in actuality. The wave function $\Psi$ is an abstract mathematical form which can be expressed in different representations, each of which is given in the formalism by different basis $\{B, B', B'', ...\}$; each basis can be interpreted as providing the set of properties which are determined. For each representation we obtain respectively $\{|\psi_{B}\rangle, |\psi_{B'}\rangle, |\psi_{B''}\rangle,...\}$.\footnote{More generally one can think in terms of density operators: firstly a $\rho$ without a definite basis, and secondly, $\{\rho_{B}, \rho_{B'}, \rho_{B''},...\}$ given by the density operator in each basis $\{B, B', B'', ...\}$.} We have to choose in which basis we are going to write the wave function (context) just like in classical mechanics we choose a certain reference frame to write our equations of motion. But in quantum mechanics, contrary to classical mechanics or special relativity, each representation/basis expresses a context which can be, in principle, \emph{incompatible} to a different context. This is where all the trouble starts: \emph{compatibility}.\footnote{For an analysis of the concept of compatibility see \cite{Aerts81} and also the very interesting passage of the book of Asher Peres \cite{Peres93}, chapter 7.}

Simon Kochen and Ernst Specker proved that in a Hilbert space $d\geq3$, it is impossible to associate numerical values, 1 or 0, with every projection operator $P_{m}$, in such a way that, if a set of {it commuting} $P_{m}$ satisfies $\sum P_{m}=\amalg$, the corresponding values of the projection operators, ${\it v}(P_{m})$, namely ${\it v}(P_{m})=0$ or $1$, also satisfy $\sum {\it v}(P_{m})=1$. This means that if we have three operators $A$, $B$ and $C$, where $[A,B] = 0$, $[A,C] = 0$ but $[B,C] \neq 0$ it is not the same to measure $A$ alone, or $A$ together with $B$, or together with $C$. The principles of quantum mechanics produce an holistic structure which is responsible for the impossibility of assigning a compatible family of truth valuations to the projection operators of different contexts. If we take ${\cal L}$ to be an orthomodular lattice and the global valuation as providing the values of all magnitudes at the same time maintaining a {\it compatibility condition}, (in the sense that whenever two magnitudes shear one or more projectors the values assigned to those projectors are the same from every context) one can state in algebraic terms the KS theorem as follows \cite{DF}:

\begin{theo}\label{CS3}
If $\mathcal{H}$ is a Hilbert space such that $dim({\cal H}) > 2$, then a global valuation, i.e. a family of compatible valuations of the contexts, over ${\mathcal L}({\mathcal H})$ is not possible. \end{theo}

\noindent This has a direct consequence in the entity-interpretation of the quantum wave function $\Psi$, because if we take an entity to be a set of definite properties which exist in actuality regardless of measuring or not, subjectivity appears as a major obstacle. In quantum theory the KS theorem shows that the concept of choice is entangled with that of existence: {\it the entity exists only because we choose}. As we have discussed before, this `subjective' or `contextual entity' is completely unacceptable in physics, a discipline which presupposes an objective account of that which is considered to be physically real. In classical mechanics, on the contrary, due to its compatible\footnote{Even though one might have incompatible experimental setups (contexts) in classical mechanics, such as those proposed by Diederik Aerts: A piece of wood which has the property of `being burnable' and of `floating' \cite{Aerts81}. One can always think of these contexts in terms of ignorance, there is no proper/ontological incompatibility, as it is always possible in principle to valuate every property without inconsistencies. It is possible to think that the piece would {\it definitely has} the mentioned properties.} (reductionistic) structure, one can neglect this level ---which we have called earlier `perspective'. Reductionistic theories do not suffer from this ``problem" because their structure always allows for a Boolean valuation. Coloring every atom in the universe (every point in phase space) would not arise a problem because the universe is nothing but the sum of these atoms. Classically, the choice of the context {\it discovers} ---rather than {\it creates}--- the elements of physical reality, which were of course already there... just like the moon is outside there regardless of our choice to look at her or not.

The perspective can be represented in infinitely many ways, each of which determines a definite relation between the properties of a system. A `new' context appears each time we choose to change the representation. The perspective cannot be {\it a priori} decomposed into elementary blocks, these holistic contexts, and the whole from which they `become', should be regarded as expressing the essential character of quantum mechanics, that of precluding the possibility of thinking about the quantum wave function in terms of the classical principles of identity, unity and totality. There is a tension between the notion of entity and the quantum formalism which has not been yet resolved. This tension is most clearly expressed through the loss of an objective account of physical reality in terms of entities (such as elementary particles).

{\smallroman
\begin{quotation}
``{\small \emph{Einstein}}'s opposition to [quantum mechanics] is
again reflected in his papers which he published, at first in
collaborations with {\small \emph{Rosen}} and {\small
\emph{Podolsky}}, and later alone, as a critique on the concept of
reality in quantum mechanics. We often discussed these questions
together, and I invariably profited very greatly even when I could
not agree with {\small \emph{Einstein}}'s view. ``Physics is after
all the description of reality" he said to me, continuing, with a
sarcastic glance in my direction ``or should I perhaps say physics
is the description of what one merely imagines?" This question
clearly shows {\small \emph{Einstein}}'s concern that the
objective character of physics might be lost through a theory of
the type of quantum mechanics, in that as a consequence of a wider
conception of the objectivity of an explanation of nature the
difference between physical reality and dream or hallucination
might become blurred." W. Pauli (\cite{Pauli94},
p.122)\end{quotation}}

As a second beat we could think that, at least $|\psi_{B}\rangle$,
the wave function in a given basis, might allow an interpretation
in terms of an entity. If we do not take a preferred basis we
normally express $|\psi_{B}\rangle$ as a linear combination of
elements, this is called a {\it superposition}. The principle of
superposition was regarded by Paul Dirac as one of the most
important features of quantum mechanics:

{\smallroman
\begin{quotation}
``The nature of the relationships which the superposition
principle requires to exist between the states of any system is of
a kind that cannot be explained in terms of familiar physical
concepts. One cannot in the classical sense picture a system being
partly in each of two states and see the equivalence of this to
the system being completely in some other state. There is an
entirely new idea involved, to which one must get accustomed and
in terms of which one must proceed to build up an exact
mathematical theory, without having any detailed classical
picture." P. Dirac (\cite{Dirac47}, p.12)\end{quotation}}

Unfortunately the idea or regarding a superposition as representing an entity does not work either. Given a superposition: $|\psi_{B}\rangle = \alpha |a\rangle + \beta |b\rangle$; if we take the elements of $|\psi_{B}\rangle$, $|a\rangle$ or $|b\rangle$ as existing both in actuality, it might happen that $|a\rangle$ and $|b\rangle$ represent a property and its opposite! So at one and the same time we have `up' and `down', `dead' and `alive', `black' and `white', of that of which we are predicating. But according to the
principle of non-contradiction, the most certain of all
principles, everything {\it is} or {\it is not} the case. If we
are to retain classical logic this leads to contradictions, and
thus to inconsistencies. One could try then, in principle, to
interpret $|\psi_{B}\rangle$ as being actually one of the two
elements of the superposition, which of them? we do not know until
we perform the experiment, which always gives one of the two
possibilities. Off course we know this idea is simply wrong, it is
incompatible with the formalism of quantum mechanics which does
not allow to provide an {\it ignorance interpretation} of the
elements of a superposition. So we got to a dead end, the elements
of the superposition do not exist in the mode of being of
actuality, the only mode of being which we are accustomed to call
`real'. But if one firmly believes that such thing as a
superposition exists, and this is what quantum mechanics tells us
if taken seriously, it seems we might be in the need of creating a
new way of dealing with these elements of the quantum structure, a
way, which should not depend on the idea of entity. We will come
back to this discussion later in section 5.2.

\subsection{The Classical Statistical Conception: Probability and Possibility}

If one learns quantum mechanics for the first time, it might seem that the theory is essentially statistical and thus refers only to probabilistic statements. One is then allowed to continue the reasoning and infer that quantum mechanics is a probabilistic theory which only talks about ensembles of systems and does not provide any direct answer refereing to an individual system. If such would be the case, there would be no interpretational problems whatsoever, $|\psi_{B}\rangle = \alpha |a\rangle + \beta |b\rangle$ would represent, not a single system, but an ensemble of them; $|\alpha|^{2}$ and $|\beta|^{2}$ being the (classical) probabilities to obtain respectively $|a\rangle$ or $|b\rangle$. It would be then possible to interpret quantum mechanics as referring to some unknown but actually existent state of affairs. This is of course the idea which the concept of probability presupposes, i.e. that there is an actual state of affairs regardless of weather we know it or not.\footnote{For a detailed discussion see \cite{DFRPPP}.} But quantum mechanics uses a non-Kolmogorovian type of probability which does not allow to interpret probability in terms of {\it ignorance} or {\it uncertainty} about an actual state of affairs. Mathematically this encounters no inconvenience whatsoever, but from a physical point of view it is a catastrophe, simply because we loose the meaning of `probability' in the quantum domain. At the time this was not easy to understand and even today, this confusion continues to burden quantum theory. This can be witnessed from many approaches which neglect the conceptual reach of the ideas which must be reformulated due to quantum mechanics. For example, according to Feynman and Hibbs, the concept of probability does not change in quantum mechanics:

{\smallroman
\begin{quotation}
``The concept of probability is not altered in quantum mechanics.
When we say the probability of a certain outcome of an experiment
is {\small {\it p}}, we mean the conventional thing, i.e.; that if
the experiment is repeated many times, one expects that the
fraction of those which give the outcome in question is roughly
{\small {\it p}}. We shall not be at all concerned with analyzing
or defining this concept in more detail; for no departure from the
concept used in classical statistics is required.

What is changed, and radically changed, is the method for
calculating probabilities. [...]" R. Feynman and A. Hibbs
(\cite{FeynmanHibbs65}, p.3)
\end{quotation}}

\noindent Feynman and Hibbs discuss the notion of probability,
relying for it, only on the relation between measurement outcomes.
The problem seems to end up by avoiding a deeper analysis of the
physical meaning of probability. On the contrary, we understand
the physical notion of probability as a gnoseological concept
which has in its heart very definite presuppositions, the most
important of which regards existence. Erwin Sch\"odinger wrote a
letter to Einstein exactly about this point many years ago:

{\smallroman
\begin{quotation}
``It seems to me that the concept of probability is terribly
mishandled these days. Probability surely has as its substance a
statement as to whether something {\small {\it is}} or {\small
{\it is not}} the case ---an uncertain statement, to be sure. But
nevertheless it has meaning only if one is indeed convinced that
the something in question quite definitely {\small {\it is}} or
{\small {\it is not}} the case. A probabilistic assertion
presupposes the full reality of its subject.'' E. Schr\"{o}dinger
(\cite{Bub97}, p.115)
\end{quotation}}

\noindent Of course the founding fathers of quantum mechanics
clearly understood the departure of quantum probability with
respect to its classical meaning, but most importantly, that this
departure precluded the possibility to maintain our classical
conception of the world.

{\smallroman
\begin{quotation}
``[...] the paper of Bohr, Kramers and Slater revealed one
essential feature of the correct interpretation of quantum theory.
{\small {\it This concept of the probability wave was something
entirely new in theoretical physics since Newton. Probability in
mathematics or in statistical mechanics means a statement about
our degree of knowledge of the actual situation.}} In throwing
dice we do not know the fine details of the motion of our hands
which determine the fall of the dice and therefore we say that the
probability for throwing a special number is just one in six. The
probability wave function of Bohr, Kramers and Slater, however,
meant more than that; it meant a tendency for something. It was a
quantitative version of the old concept of `potentia' in
Aristotelian philosophy. It introduced something standing in the
middle between the idea of an event ant the actual event, a
strange kind of physical reality just in the middle between
possibility and reality.'' W. Heisenberg (\cite{Heisenberg58},
p.42, emphasis added)
\end{quotation}}

In relation to the impossibility of providing an ignorance interpretation of the elements present in the quantum formalism I would also like to call the attention to the well known distinction provided by Bernard D'Espagnat in his book, {\it Conceptual Foundations of Quantum Mechanics} \cite{D'Espagnat76}, between {\it proper} and {\it improper} mixtures. D'Espagnat shows that a mixture which is obtained from a pure quantum state cannot be interpreted in terms of ignorance without getting into contradictions. The mixtures which one obtains by tracing degrees of freedom from an original pure state, even though present
exactly the same formal structure than a ``common", ``proper" mixture, cannot be supplemented with an ignorance interpretation
of its terms. D'Espagnat calls these quantum mixtures {\it improper}, in contraposition to classical mixtures which are
called {\it proper}. I regard this as one of the most important conceptual developments of the last decades, its importance has to
do with the possibility of understanding better, on the one hand, what we do  mean when we talk about a `quantum mixture', and on the
other hand, the impossibilities of the quantum formalism to provide an account of an actual state of affairs.

However, not only the notion of probability, but also that of {\it possibility} finds within the formal structure of quantum theory serious inconveniences to relate itself closely to our classical understanding. In \cite{DFRAnnalen} and \cite{DFRIJTP}, together with Graciela Domenech and Hector Freytes we applied algebraic and topological tools in order to study the structure of the orthomodular lattice of actual propositions enriched with modal propositions. In this work we developed the following frame: If ${\cal L}$ is an orthomodular lattice and ${\cal L}^{\diamond}$ a Boolean saturated orthomodular one such that ${\cal L}$ can be embedded in ${\cal L}^{\diamond}$, we say that ${\cal L}^{\diamond}$ is a modal extension of ${\cal L}$. Given  ${\cal L}$ and a modal extension ${\cal L}^{\diamond}$,  we define the {\it possibility space} as the subalgebra of ${\cal L}^{\diamond}$ generated by $ \{\diamond P : P \in {\cal L} \} $. We denote by $\diamond {\cal L}$ this space and it may be proved that it is a Boolean subalgebra of the modal extension. The possibility space represents the modal content added to the discourse about properties of the system. We may define a {\it global valuation} over ${\mathcal L}({\mathcal H})$ as the family of Boolean homomorphisms $(v_i: W_i \rightarrow {\bf 2})_{i\in I}$ such that $v_i\mid W_i \cap W_j = v_j\mid W_i \cap W_j$ for each $i,j \in I$, being $(W_i)_{i\in I}$ the family of Boolean sublattices of ${\mathcal L}({\mathcal H})$. This global valuation would give the values of all magnitudes at the same time maintaining a {\it compatibility condition} in the sense that whenever two magnitudes shear one or more projectors, the values assigned to those projectors are the same from every context. Within this frame, the actualization of a possible property acquires a rigorous meaning. If $f: \diamond {\cal L} \rightarrow {\bf 2}$ is a Boolean homomorphism, an actualization compatible with  $f$ is a global valuation $(v_i: W_i \rightarrow {\bf 2})_{i\in I}$ such that $v_i\mid W_i \cap \diamond {\cal L} = f\mid W_i \cap \diamond {\cal L} $ for each $i\in I$. Compatible actualizations represent the (logical) passage from possibility to actuality. When taking into account compatible actualizations from different contexts, the following KS theorem for modalities can be proved \cite{DFRAnnalen}:

\begin{theo}\label{MKS}
Let $\cal L$ be an orthomodular lattice. Then $\cal L$ admits a global valuation iff for each possibility space there exists a Boolean homomorphism  $f: \diamond {\cal L} \rightarrow {\bf 2}$ that admits  a compatible actualization.
\end{theo}

The modal KS (MKS) theorem shows that no enrichment of the orthomodular lattice with modal propositions allows to circumvent the contextual character of the quantum language. As it has been discussed in \cite{DHR07} a further conclusion which can be derived from the MKS theorem is that the formalism of quantum mechanics does not only deny the possibility of talking about an `actual entity', but even the term `possible entity' remains a meaningless notion within its domain of discourse.

\subsection{Dirac's Problematic Definition: `State of a System'}

Dirac was obviously aware of the limitations imposed by quantum
mechanics to our classical picture of the world in terms of
entities, this becomes evident from the preface to the first
edition of his book, {\it The Principles of Quantum Mechanics}:

{\smallroman
\begin{quotation}
``The methods of progress in theoretical physics have undergone a
vast change during the present century. The classical tradition
has been to consider the world to be an association of observable
objects (particles, fluids, fields, etc.) moving about according
to definite laws of force, so that one could form a mental picture
in space and time of the whole scheme. This led to a physicist
whose aim was to make assumptions about the mechanism and forces
connecting these observable objects, to account for their
behaviour in the simplest possible way. It has become increasingly
evident in recent times, however, that nature works on a different
plan. Her fundamental laws do not govern the world as it appears
in our mental picture in any very direct way, but instead they
control a substratum of which we cannot form a mental picture
without introducing irrelevancies." P. Dirac (29 May 1930,
\cite{Dirac47}, preface to the first edition)\end{quotation}}

In the second edition of his book, Paul Dirac continues to analyze
the relation between quantum mechanics and classical concepts,
coming to attack in this opportunity the notion of `state' and its
possible meaning in the quantum formalism:\footnote{I wish to
thank Carlo Rovelli for pointing out this important passage to
me.}

{\smallroman
\begin{quotation}
``The main change [in the book] has been brought about by the use
of the word `state' in a three-dimensional non-relativistic sense.
It would seem at first sight a pity to build up the theory largely
on the basis of non-relativistic concepts. {\small{\it The use of
the non-relativistic meaning of `state', however, contributes so
essentially to the possibilities of clear exposition as to lead
one to suspect that the fundamental ideas of the present quantum
mechanics are in need of serious alteration at just this point}},
and that an improved theory would agree more closely with the
development here given than with a development which aims at
preserving the relativistic meaning of `state' throughout." P.
Dirac (27 November 1934, \cite{Dirac47}, preface to the second
edition)\end{quotation}}

At this point we are only interested in making clear the
irrelevancy of the notion of `state of the system' in the scheme provided by
quantum mechanics. David Finkelstein is someone who has also
called the attention of the fact that it would be better to do
without this notion.

{\smallroman
\begin{quotation}
``One is liable to think that `the state of the system' is an
indispensable element of the quantum theory, simply because it is
found in many expositions. Even the founding fathers, who knew
better, occasionally lapsed into phrases like `the state of the
system,' though in contexts that made it clear that they did not
attribute physical reality to the construct. To make it explicit
that there is no longer room or need for this construct in quantum
physics, we review here a formulation which avoids it from the
start. All of this is at least implicit and often explicit in the
writings of Heisenberg and Bohr." D. Fikelstein
(\cite{Finkelstein05}, p.2) \end{quotation}}

What is important to notice is that the notion of `state' goes
together with the idea of `object', it has only meaning when
presupposing the existence of something which has such `state',
that is why one talks about `the state {\it of a system}'. But, as
we saw above, there is no absolute state of a system in quantum
mechanics, the state of the system exists (in actuality) only when
the choice of the context has taken place. Most importantly, the KS theorem makes clear the fact that it is not possible
to think of this choice as revealing a preexistent (actual)
reality.

In order to find a  way out, there are some approaches which point
out that by relativizing concepts one might be able to
successfully interpret quantum mechanics. Such is the case of
Carlo Rovelli who presented the idea that one should reject the
notion of: {\it absolute state} or {\it observer independent state
of a system} (observer-independent values of physical quantities).
In his \emph{Relational Interpretation of Quantum
Mechanics}\footnote{This interpretation was exposed in
\cite{Rovelli96}.} this notion is replaced in favor of: {\it state
relative to something}. Rovelli argues that the conclusion derives
from the observation that the experimental evidence at the basis
of quantum mechanics forces us to accept that distinct observers
give different descriptions of the same events.\footnote{This
interpretation of Rovelli has been discussed in detail in
\cite{deRondeRel}.} Also David Finkelstein points out that one
could advance in this direction.

{\smallroman
\begin{quotation}
``Quantum theory was consciously and explicitly modelled on
special relativity. The theory was formulated operationally to
free it of certain idols. The role that the Lorentz group plays in
special relativity is played by the unitary group in Dirac's
transformation theory of quantum theory. Quantum theory
relativized the construct of `the state of the system,' implicitly
absolute, and replaced it by `the state of the system relative to
this experimental frame.' Nevertheless the construct of state was
still useful and is still used.[...]

The quantum relativity of the state seems to violate common sense
even more than the classical relativity of time, though it seems
to agree well with experiment. Each person is pretty well-steeped
in both commutative logic and absolute time by the time he or she
encounters quantum theory and special relativity. Every generation
will have to go through these processes of relativizing the
concepts of time and state and of who knows what else to come." D.
Finkelstein (\cite{Finkelstein05}, p.2)
\end{quotation}}

\noindent We certainly agree that in some cases the idea of
relativizing concepts has been of great help to find a way out,
and indeed this has worked out very well in the history of
physics. It is also true that by relativizing the concepts of
space and time Einstein was able to produce a new conception of physical reality. However, the point we would like to make clear is that one cannot relativize the idea of entity in the way it is done in
quantum mechanics and go away with it, because, if we do so,
everything which is left behind looses its meaning.

The similarities between quantum mechanics and relativity theory
are certainly interesting but the most important thing to learn
abut their relation is their difference, the distance between
them. Even Niels Bohr discussed the common features of both
relativity and quantum mechanics trying to find a valuable
analogy which would help understanding quantum theory. As
commented by Max Jammer, in 1929 Bohr compared in three different
aspects his approach in quantum mechanics with Einstein's theory
of relativity:

{\smallroman
\begin{quotation}
``[...] Concerning the first two points of comparison Bohr was
certainly right. But as to the third point of comparison, based on
the assertion that relativity theory reveals `the subjective
character of all concepts of classical physics' or, as Bohr
declared again in the fall of 1929 in an address in Copenhagen,
that `the theory of relativity remind us of the subjective
character of all physical phenomena, a character which depends
essentially upon the motion of the observer,' [...] {\small{\it
Bohr overlooked that the theory of relativity is also a theory of
invariants and that, above all, its notion of `events,' such as
the collision of two particles, denotes something absolute,
entirely independent of the reference frame of the observer and
hence logically prior to the assignment of metrical attributes.}}"
M. Jammer (\cite{Jammer74}, p.132, emphasis added)\end{quotation}}

\noindent In special relativity one can still talk about an actual
reality, there are {\it events} and these events can be interpreted as existing regardless of
being or not being observed. In special relativity the mathematical structure which relates events in space-time allows (through the invariants present in the theory) to retain
an objective picture of physical reality. However, this is not the case
when we relativize the notion of state of a system in quantum
mechanics, because by doing this, we are relativizing the very
notion of physical reality. We have argued that this cannot be accepted in
physics whose basic presupposition is that something like physical reality,
call it Nature, exists. By relativizing the idea of state of a
system not only the idea of system losses its meaning but even the
the notion of physics losses its content.

Even though we agree with Carlo Rovelli and David Finkelstein in
the important point that the notion of `state of a system' is
superfluous in quantum mechanics, we disagree in what should be
done in this respect. In this point we propose a radical move, the
extreme consequence we need to derive from this problem is not to
relativize, but to forget completely about the notion of `state of
the system'. The definition provided by Dirac is the seed of the
interpretational problems which will be later evidenced in
different levels. As recalled by Rovelli himself: ``Heisenberg's
insistence on the fact that the lesson to be taken from the atomic
experiments is that we should stop thinking of the `state of the
system', has been obscured by the subsequent terse definition of
the theory in terms of states given by Dirac."\footnote{Quoted
from \cite{Rovelli96}, p.19.} Talking about a state of a system is
nothing but presupposing that quantum mechanics talks about
entities. The idea that there are such systems (entities) reminds
us of the Socratic questioning upon which one is already trapped
if one tries to answer from the structure delivered by the
question itself. It is the presupposed structure of the question
which limits the possibilities of providing an answer. If one
tries to answer the question from within the hidden structure one
gets into a labyrinth, and hunted by the Minotaur one is sure to
dye inside. But we, we have the thread of Ariadna (our classical
language), a thread which we should not overestimate (as a
fundamental path). We must remain suspicious because scientist we
are. We must be cautious as the thread might constitute the
labyrinth itself. The thread might hide the possibilities to
create new paths, a new way of understanding the problem. We have
to use the thread not to get out, just by pulling in the one side,
but rather by using it to view the distinctions, the differences,
and in this way \emph{understand} the labyrinth. We have to escape
not by finding a secrete path hidden in the formalism but rather
by elevating ourselves by means of abstraction. We have to {\it
create} a way out.

\section{The Concept of `Entity' as an Epistemological
Obstruction}

Even though we recognize its importance in occidental thought, we
believe that the idea of entity appears, in the context of quantum
mechanics as, what Gaston de Bachelard \cite{Bachelard} calls, an
{\it epistemological obstruction}; i.e. an idea which restricts
our possibilities to imagine the physics provided by quantum
mechanics. Quantum mechanics is the first physical theory which
provides a clear example of the impossibility of interpreting its
formalism subsumed by the idea of entity. Maybe this is why von
Weizs\"acker wrote that ``[in quantum theory] the dissolution of
the traditional concepts such as space, time, matter,
determination, produces in every man which seriously confronts it,
in the first place, the feeling of being confronted with
nothingness."\footnote{Quoted from \cite{vW74}, p. 247, our
translation.} Quantum mechanics stands in the limits of an abysm,
the radicalness of its conceptual break through is still today not
completely acknowledged. Quantum theory tackles once and again the
conception of physical entity, and with it, all of our classical
worldview. Our language is entangled with our classical conception
of the world, a world which is expressed in terms of subject,
object, predicate, etc. It is this same language which exposes the
limits of our world. Quantum mechanics stands beyond this specific
conceptualization, waiting for a new language which can express
its power.

In order to make experiments we need the idea of `objects'
(particles, apparatus, photographic plates, etc.), so in a
particular sense the connection between quantum mechanics and
classical physics seems quite direct. The paradox appears when we
do not recognize that even though we use a classical apparatus
{\it the experiments that we perform when considering elementary
particles are not part of our classical experience}. The
presuppositions we make, already by using the quantum formalism,
go completely against the idea of physical reality as constituted
by entities (particles, apparatus, photographic plates, etc.). It
is this conjunction which cannot be resolved. Each principle of
Aristotelian logic which is at the basis of our understanding
becomes rotten in the quantum formalism. As a matter of fact, all
the discussions regarding the interpretation of quantum mechanics
explicitly or implicitly refer to the impossibility of thinking in
classical terms. Quantum mechanics deconstructs the conceptual
scheme provided by classical physics and with it, the particles,
the apparatuses, the photographic plates and whatever object that
we might use to perform a measurement.

The structure of thought which we use in classical physics is that
guided by classical Aristotelian logic and its principles,
however, the formalism of quantum mechanics and its experience
seems to contradict each one of them. Quantum mechanics places new
principles such as {\it indetermination}, {\it complementarity} and
{\it superposition}. An important point regards the interpretation of
such principles in terms of, either providing `consistent
knowledge of classical experience' or, as providing `the structure
of thought of a completely new experience', that one expressed by
quantum theory.

\subsection{Indetermination Instead of Existence}

Werner Heisenberg \cite{Heisenberg27} presented in 1927 one of the
most important papers of the 20th century. In this paper called:
{\it Uber den anschaulichen Inhalt der quantentheoretischen
Kinematik und Mechanic}, the principle of existence and its
direct relation to actuality was attacked through a set of
indeterminacy relations. As a first characterization we might say
that the {\it principle of indetermination} expresses the
impossibility of assigning exact simultaneous values to the
position and momentum of a particle. According to Heisenberg the
properties are not determined until the measurement has taken
place, there is no actual state of affairs related to the evolution of the quantum wave
function, or in his own words: ``I believe that one can formulate the
emergence of the classical `path' of a particle pregnantly as
follows: {\it the `path' comes into being only because we observe
it.}". As recalled by Heisenberg himself it was Einstein's
recommendation which guided him:

{\smallroman
\begin{quotation}
``[In the transformation theory by Dirac and Jordan] one could
transform from {\small $\psi (q)$} to {\small $\psi (p)$}, and it
was natural to assume that the square {\small $|\psi (p)|^{2}$}
would be the probability to find the electron with momentum p. So
gradually one acquired the notion that the square of the wave
function, which by the way was not the wave function in
three-dimensional space but in configuration space, meant the
probability for something. With this knowledge we returned to the
electron in the cloud chamber. Could it be that we had asked the
wrong question? I remember Einstein telling me, `it is always the
theory which decides what can be observed.' And that meant, if it
was taken seriously, that we should not ask: `How can we represent
the path of the electron in the cloud chamber?' We should ask
instead: `Is it not perhaps true that in nature only such
situations occur which can be represented in quantum mechanics or
wave mechanics?" W. Heisenberg (\cite{Heisenberg73}, p.269)
\end{quotation}}

\noindent If taken to its last consequences, Einstein's
recommendation means that `the theory' expresses the conditions of
possibility to determine what is to be considered `experience'.
Our conception of reality is modeled in this way by the theory
itself which determines the ontological and epistemological
conditions over which it provides `meaning'. It is the theory
which determines the limits of what is to be considered experience
and physical reality. If we accept that ``there are no facts without
a theory'', in quantum mechanics it is meaningless to say
that the quantum system, conceived in terms of the the wave function, possesses a set of definite properties. Properties which exist (in actuality) regardless of weather we
observe them or not. Heisenberg's principle, if taken seriously, this is, in
terms of an {\it ontological interpretation}, has nothing to do
with ignorance. Unfortunately, Bohr's pressure to subsume
this principle within his own complementary scheme forced
the subsequent {\it gnoseological discussions} in terms of
experimental impossibilities \cite{GdeR07}.

Even though Heisenberg had started by analyzing experiments, after
having found a consistent way of recovering ``the observed"
through his mathematical scheme of matrix mechanics, he was
stopped from going further and taking this same principle as a
guiding line to determine future experience. Heisenberg, returned
in his footsteps and remained within the limits imposed by
classicality. Instead of taking his principle along the
ontological road of Einstein, Heisenberg followed Bohr's
gnoseological path. Such trip had no other goal than to justify quantum
theory from the heights of classical thought. The pressure of Bohr
can be read in the ``Addition in Proof" to Heisenberg's
foundational paper:

{\smallroman
\begin{quotation}
``After the conclusion of the forgoing paper, more recent
investigations of Bohr have led to a point of view which permits
an essential deepening and sharpening of the analysis of
quantum-mechanical correlations attempted in this work. In this
connection Bohr has brought to my attention that I have overlooked
essential points in the course of several discussions in this
paper. Above all, the uncertainty in our observation does not
arise exclusively from the occurrence of discontinuities, but is
tied directly to the demand that we ascribe equal validity to the
quite different experiments which show up in corpuscular theory in
the one hand, and in the wave theory in the other hand. [...] I
owe great thanks to Professor Bohr for sharing with me at an early
stage the results of these more recent investigations of his-to
appear soon in a paper on the conceptual structure of quantum
theory- and for discussing them with me." W. Heisenberg
(\cite{Heisenberg27} quoted from \cite{WZ}, p.83)
\end{quotation}}

Niels Bohr considered the wave-particle duality present in the
double slit experiment (section 6.1) as expressing the most
important character of quantum theory. What Bohr had in mind, as
we shall see later, was to resolve this duality through the
principle of complementarity. Bohr's agenda was focused in
fulfilling the consistency requirements of the quantum formalism
to apply the well known classical scheme, the discussions which
followed took Heisenberg's principle only as providing the limits
of certainty. The classical scheme would then remain that which
secured the knowledge provided by quantum theory, and analogously,
Heisenberg's {\it uncertainty relations} that which secured the
knowledge provided by the more general principle of
complementarity.

{\smallroman
\begin{quotation}
``Bohr wanted to pursue the epistemological analysis one step
further [than Heisenberg], and in particular to understand the
logical nature of the mutual exclusion of the aspects opposed in
the particle-wave dualism. From this point of view the
indeterminacy relations appear in a new light. [...] The
indeterminacy relations are therefore essential to ensure the
consistency of the theory, by assigning the limits within which
the use of classical concepts belonging to the two extreme
pictures may be applied without contradiction. For this novel
logical relationship, which called in Bohr's mind echoes of his
philosophical meditations over the duality of our mental activity,
he proposed the name `complementarity', conscious that he was here
breaking new ground in epistemology." L. Rosenfeld (\cite{WZ}, p.59)\end{quotation}}

Pekka Lahti proved in his thesis \cite{Lahti80}
that Heisenberg's principle is logically independent of Bohr's
principle of complementarity. Today, we have more elements to make precise the relation between
these principles. Firstly, it is important to notice
that Heisenberg's relations can be derived directly from the
mathematical scheme of the theory, as a direct consequence of the
quantum postulate. At first sight it might seem that the denial of
the existence of properties which are not ``observed" has an
operational ground, and this might have been the case, however, we
believe the most important consequence can be derived if this move
is read from somewhat different angle. The droplets in the cloud
chamber show that that which we observe \emph{appears} and
\emph{disappears}. When one sees something one is accustomed to
say that that which is observed is ``actually (in reality) there",
but what is the mode of being of something which disappears, of
something which is not present in actuality? This is the problem
which Heisenberg encountered. If taken through the lines of
thought of Einstein himself, acknowledging that in physics ``there
is no difference between observable and non-observable",
Heisenberg's principle appears in a completely new light,
referring to a different mode of existence to that of actuality.

However, following Bohr's recommendation, Heisenberg's principle was interpreted in terms of ignorance, as
{\it uncertainty}, and even explained through a set of `gedankenexperiments' which were expressions of an {\it experimental
impossibility} ---in contraposition to an analysis over the conditions which make possible the form of experience demanded by
the theory. But as noted by Jaan Hilgevoord and Joos Uffink: ``[...]it is remarkable that in his later years Heisenberg put a somewhat different gloss on his relations. In his autobiography {\it Der Teil und das Ganze} of 1969 he described how he had found his relations inspired by a remark by Einstein that `it is the theory which decides what one can observe' ---thus giving precedence to theory above experience, rather than the other way around." Most interestingly for us is the fact that ``Some years later he even admitted that his famous discussions of thought experiments were actually trivial since `[...] if the process of observation itself is subject to the laws of quantum theory, it must be possible to represent its result in the mathematical scheme of this theory'."\footnote{Quoted from
\cite{HU}.}

To take quantum mechanics seriously is to believe that quantum
mechanics expresses some feature of reality and not only a
consistent discourse which allows us to analyze experiments
expressed in terms of classical mechanics. The algorithmic
conception of quantum mechanics as providing results of
measurement outcomes goes completely against the very idea of
doing physics. At this point we choose to remain close to the
meaning provided by the principle itself. If we think that quantum
mechanics is telling something about the world, if we think that
it expresses an objective account of physical reality, we are then
forced to understand Heisenberg's principle in terms of a mode of
being, in terms of {\it indetermination}.

The {\it indetermination principle} lyes parallel to the {\it
principle of existence} (of classical logic) but stating something
completely different. The principle of indetermination refers to
the mode of existence of properties in quantum mechanics, it
states that the properties of a quantum system remain
indetermined, in the potential form of the being. Potentiality and
indetermination are concepts which stand side by side, just like
actuality and determination. If we regard quantum theory as saying
something about the world, Heisenberg's principle should remain an
ontological presupposition for experience as expressed by quantum
theory, this is why we have stated several times in the past that
{\it quantum mechanics creates a new experience}.

\subsection{Complementarity Instead of Non-Contradiction}

At the same time that Heisenberg had produced his indetermination
principle, Niels Bohr appeared in the scene with a principle of
his own: {\it the principle of complementarity}. It is, at least, not
completely obvious what Bohr meant with this principle:

{\smallroman
\begin{quotation}
``Complementarity is no system, no doctrine with ready-made
precepts. There is no via regia to it; no formal definition of it
can be found in Bohr's writings, and this worries many people.
[...] Bohr was content to teach by example. He often evoked the
thinkers of the past who had intuitively recognized dialectical
aspects of existence and endeavored to give them poetical or
philosophical expression." L. Rosenfeld (\cite{WZ},
p.85).\end{quotation}}

\noindent Even though we regard this indefinite exposure as the
richness itself of Bohr's discourse, for our immediate proposes we
will discuss the possibility of limiting the meaning of
complementarity.\footnote{As noted by Pekka Lahti in
\cite{Lahti80}, p.801, ``In reading Bohr's writings one may easily
form the impression that the notion of complementarity does appear
in many different connections. However, one can distinguish
between four categories of statements which cover most uses of
this notion. These are the following: (a) complementarity as a
relation between descriptions. like space-time description and
causal description, (b) complementarity as a relationship between
elementary physical concepts, like position and momentum, (c)
complementarity of the particle picture and the wave picture, and
(d) complementarity as a relationship between phenomena demanding
mutually exclusive experimental arrangements. [...] It seems to us
that in developing his viewpoint of complementarity Bohr gradually
shifted the emphasis from category (a) to category (b), and
ultimately `unified' the first three seemingly different notions
of complementarity under the appearing in the category (d)."}
Bohr's main discussion related directly to the problem of
objectivity, complementarity was meant here as a general {\it
regulative principle} which would allow to discuss consistently
mutually incompatible experimental arrangements. Bohr's starting
point was the wave-particle duality and the idea that: ``We must,
in general, be prepared to accept the fact that a complete
elucidation of {\it one and the same object} may require diverse
points of view which defy a unique description."\footnote{Quoted
from \cite{Bohr29}, emphasis added.} For the description of
certain atomic phenomena we need a `particle picture' while for
others we need a `wave picture', using both pictures
simultaneously leads to contradictions. According to Bohr, it is
the idea of complementarity, as a regulative principle, which
allows to secure the consistency of knowledge and to recover an
objective description of physical reality.

{\smallroman
\begin{quotation}
``On the lines of objective description [I advocate using] the
word {\small{\it phenomenon}} to refer only to observations under
circumstances whose description includes an account of the whole
experimental arrangement. In such terminology, the observational
problem in quantum physics is deprived of any special intricacy
and we are, moreover, directly reminded that every atomic
phenomenon is closed in the sense that its observation is based on
registrations obtained by means of suitable amplification devices
with irreversible functioning such as, for example, permanent
marks on a photographic plate, caused by the penetration of
electrons into the emulsion. In this connection, it is important
to realize that the quantum-mechanical formalism permits well
defined applications referring only to such closed phenomena." N.
Bohr (\cite{WZ}, p.3)
\end{quotation}}

The definition of phenomenon relied for Bohr in the use of {\it
classical language}, this was the limit which even quantum
mechanics had to respect. Bohr sustained the idea that: ``it would
be a misconception to believe that the difficulties of the atomic
theory may be evaded by eventually replacing the concepts of
classical physics by new conceptual forms."\footnote{Quoted from
\cite{WZ}, p.7.} For us, complementarity, rather than unifying,
expresses the fact that one cannot put together incompatible
experience of one and the same object through classical ideas. But
then, we come back to our point of departure, what does it mean to
have {\it one and the same object}?

Complementarity circumvents the principle of non-contradiction,
but makes explicit at the same time its exclusion from classical
logic. Its relation to paraconsistent logics has been discussed by
Newton da Costa and D\`ecio Krause in their article {\it The logic of
complementarity} (see also \cite{DallaChiaraGiuntini}).

{\smallroman
\begin{quotation}
``[...] it is perfectly reasonable to regard complementary aspects as
\small{{\it incompatible}}, in the sense that their combination into a
single description may lead to difficulties. But in a theory
grounded on standard logic, the conjunction of two theses is also
a thesis; in other words, if \small{ $\alpha$} and \small{ $\beta$} are both theses or theorems of a theory (founded on
classical logic), then \small{ $\alpha \wedge \beta$} is also a
thesis (or a theorem) of that theory. This is what we intuitively
mean when we say that, on the grounds of classical logic, a 'true'
proposition cannot 'exclude' another 'true' proposition. In this
sense, the quantum world is rather distinct from the 'classical',
for although complementary propositions are to be regarded as
acceptable, their conjunction seems to be not." N. da Costa and D.
Krause (\cite{CostaKrause03}, p.5)
\end{quotation}}

Even though it is clear that complementarity stands outside the
limits imposed by classical thought, a main point of discussion
regards its relation to classical ideas, and thus, the question of
what is a proper interpretation of this principle? Niels Bohr's
ideas were focused in respecting the basic pillars of classical
thought. Through his {\it gnoselogical interpretation} the
complementarity principle was understood as providing the
constrains for a consistent classical discourse, being applied to
the relation between classical representations. Contrary to this
idea, we propose to consider the principle of complementarity from
an {\it ontological stance}, not simply relating classical
schemes, but rather as providing the logical constitution of the
relation between quantum-properties. This deeper interpretation of
the principle determines a new reality, as expressed by quantum
mechanics, independent of classical physics. We will come back to
this in section 5.1.

\subsection{Superposition Instead of Identity}

The concept of identity sinked as well in the waves of the quantum
formalism. Erwin Schr\"odinger was very clear about
this departure:

{\smallroman
\begin{quotation}
``I mean this: that the elementary particle is not an individual;
it cannot be identified, it lacks `sameness.' [...] The
implication, far from obvious, is that the unsuspected epithet
`this' is not quite properly applicable to, say, an electron,
except with caution, in a restricted sense, and sometimes not at
all." E. Schr\"odinger (\cite{Sch98}, p.197)
\end{quotation}}

However, even though Schr\"odinger was radical enough to proclaim the loss of identity in quantum mechanics, he remained within the entity conception of thought. He clearly understood that the notion of identity was left aside, but he would still remain within the linguistic structure determined by the notion of `elementary particle' ---constituting another paradox in relation to classical thought--- providing meaning to the concept of `entity' with no `identity'.\footnote{This will be discussed more closely in \cite{deRondeNoEntity}, see also \cite{KrauseEntity}.}

The notion of identity in quantum mechanics has been discussed mainly in relation to the problem of indistinguishable particles, as it is well known the way in which we count elementary particles is not classical, permutations of particles are not taken into account and thus, the statistics change drastically. However, we believe that this is only a very specific aspect of a larger problem which can be envisaged from different angles \cite{DHR07}. Another way of looking at the problem of identity is through the notion of superposition, whose  mathematical structure cannot be subsumed into the notion of `identity'. From a classical viewpoint it is not possible to bring together something and its opposite, it makes no sense to talk about an identity which possesses contraries. A superposition reflects one of the strangest characters of the quantum, presenting clear constrains to a classical interpretation. We still have to answer the question: what is a superposition? We will come back to this problem later in section 5.2 and 6.2.

\subsection{Ontology Instead of Gnoseology}

So why should we talk about entities if every single principle
which structures this idea seems to vanish in the quantum
formalism? The answer is simple: our language is bounded by this
same structure; and like Bohr used to say many times: ``we are
suspended in language". However, contrary to Bohr who stated that
no conceptual development would help us in solving the problems
into which quantum mechanics confronts us, we think that the
development of new thought-forms can certainly provide the missing
piece of the puzzle; i.e. a complete elucidation of the meaning of
quantum theory \cite{deRondeUQMCD}.

The principle of indetermination (instead of the principle of existence), the
principle of complementarity (instead of the principle of
non-contradiction) and the principle of superposition (instead of
the principle of identity) should configure this new thought
forms, which must later turn into a complete language. It is
through this new language that we must recover an ontological
account of quantum mechanics. In order to go further we must go back to Einstein's ontological
concern. If quantum mechanics is to be understood as providing a
picture of physical reality we must avoid Bohr's gnoseological
trap and continue to interpret each principle of quantum mechanics
as giving us access to the real.

In the following diagram we present a short review of the related concepts:\\

\begin{tabular}{|c|c|c|}

\hline
& \textbf{Gnoseological Interpretation} & \textbf{Ontological Interpretation} \\
\hline
\textsl{Heisenberg's} & \footnotesize{{\it Regulates complementarity.}} & \footnotesize{{\it Constitutive principle.}}\\
\textsl{relations/principle} & \footnotesize{{\it Constrains of knowledge}} & \footnotesize{{\it Determines the mode of existence}}\\
\textsl{Bohr's} & \footnotesize{{\it about properties.}} & \footnotesize{{\it of q-properties.}}\\
\textsl{} & \footnotesize{{\it UNCERTAINTY RELATIONS}} & \footnotesize{{\it PRINCIPLE OF INDETERMINATION}}\\
\hline
\textsl{Bohr's} & \footnotesize{{\it Regulative principle.}} & \footnotesize{{\it Constitutive principle.}}\\
\textsl{relations/principle} & \footnotesize{{\it How classical representations relate.}} & \footnotesize{{\it How q-properties relate.}}\\
\textsl{} & \footnotesize{{\it COMPLEMENTARITY RELATIONS}} & \footnotesize{{\it PRINCIPLE OF COMPLEMENTARITY}}\\
\hline
\textsl{Superposition} & \footnotesize{{\it Mathematical algorithmic device.}} & \footnotesize{{\it Constitutive principle.}}\\
\textsl{state/principle} & \footnotesize{{\it No image.}} & \footnotesize{{\it Explained in terms of faculties} (sec. 5.2).}\\
\textsl{} & \footnotesize{{\it MATHEMATICAL STATE}} & \footnotesize{{\it PRINCIPLE OF SUPERPOSITION}}\\
\hline
\end{tabular}
\\
\\

We believe that, as it stands, quantum theory still makes reference directly or indirectly to entities, but then we must acknowledge ---from a direct analysis of the mathematical formulation of the theory--- that this entities are created through our choices. Quantum mechanics, if talking about entities, is closer to a theory which describes what we imagine, and oops... that which we imagine turns out to be reality!

We need to develop a new ontology which can bring into stage that of which quantum mechanics is talking about, the understanding of the principles in terms of providing the ontological background of the theory needs to be reconsidered.

\section{Actuality vs. Potentiality or Classical vs. Quantum}

The first philosophers believed in the existence of {\it physis},
contrary to the Sophists who believed in the laws of man and the
\emph{polis}, these so called physicists, placed the fundament of
thought in Nature. The most important problem physicists had to
deal with was that exposed by two pre-Socratic thinkers known by
the names of Heraclitus and Parmenides, roughly speaking: what is
movement?

The received view presents these pre-Socratic thinkers as
approaching the problem from two, seemingly opposed positions.
Hercalitus of Elea, stated the theory of flux, a doctrine of
permanent motion and unstability in the world. The consequences of
this doctrine were, as both Plato and Aristotle stressed
repeatedly, the impossibility to develop stable, certain knowledge
about the world, for an object, changing each instant, does not
allow for even to be named with certainty, let alone to be
`known', i.e., assigned fixed, objective characteristics.
Parmenides was placed at the opposite side, teaching the
non-existence of motion and change in reality, reality being
absolutely One, and being absolutely Being.\footnote{Contrary to the orthodox view, one could state however following K. Verelst and B. Coecke \cite{VerelstCoecke}, that: ``[...] the `contradiction' seen by classical philosophy between Heraclitus and Parmenides is not necessarily a correct understanding of the earlier `philosophies'. One could as well infer that Heraclitus and Parmenides do articulate the same world-experience, the former as the experience of reality over a lapse of time, the latter as the experience of the absolute reality of this moment.''} Aristotle solved the problem by presupposing a certain stability
of the Being, structuring a set of principles, as those which
governed thought and Nature. The principles of {\it existence},
{\it identity} and {\it non-contradiction}, constitute the basic
structure of the idea of entity as that of which reality is
constituted in {\it actuality}. However, the structure created by
these logical and ontological principles was completely statical,
no movent or becoming could arise from it, this is why Aristotle
was in need of God, a {\it first mover} which he characterized as
being in {\it pure acto}. The importance of the concept of
potentiality, which was first placed by Aristotle in equal footing
to actuality, was soon diminished. The choice to conceive the
immobile motor as {\it pure acto} determined the fate of western
thought through the path of actuality. Potentiality became mere
possibility, and thought only in relation to the latter, its
conception, as a different mode of the being, was soon forgotten.

\subsection{Classical Mechanics as a Theory of the Actual}

The idea of regarding actuality as the real continued to rule not
only in philosophy but also in physics. However, it was only through the
development of the {\it continuous} by Leibnitz and Newton, that
it was possible to extend the physical conception of actuality
into a closed mathematical formulation.
Classical mechanics is the final stage of a complete theory which
studies entities which exist in the mode of being of actuality.
Within this description everything becomes determined and actual.
The statement of Parmenides, can be extended in time, and that
which \emph{is} will remain, that which \emph{is not} will never
be. Classical Newtonian mechanics is the final stage of the long
trip initiated by Plato and Aristotle, the mathematical structure
of the {\it Principia} the actual story of the world, physics, the
theory of actuality.

In the 20th century the socratic questioning remained still
present through the structure of classical thought which made
impossible to express anything which was not the case, which was
not actual. Even Heisenberg, who was the first to propose to think
in terms of `potentia' remained prisoner of the old idol, and
phrases like ``strange kind of physical reality", ``vague
connection with reality'' or ``not as real" accompanied the
negative characterization of potentiality always thought in terms
of actuality. For the founding fathers of quantum mechanics, the
idea of actuality remained a ghost, which appeared and reappeared
each time they looked away from her site. As self reminders of the
history of western thought, one can find through the passages of
their writings, long shadows of actuality which continue to our
days.

{\smallroman
\begin{quotation}
``Reality resists imitation trough a model. So one lets go of
naive realism and leans directly on the indubitable proposition
that {\small {\it actually}} (for the physicist) after all is said
and done there is only observation, measurement. Then all our
physical thinking thenceforth has the sole basis and as sole
object the results of measurements which can in principle be
carried out, for we must now explicitly {\small {\it  not}} relate
our thinking any longer to any kind of reality or to a model. All
numbers arising in our physical calculations must be interpreted
as measurement results. But since we didn't just now come into the
world and start to build up our science from scratch, but rather
have in use a quite definite scheme of calculation, from which in
view of the great progress in Q.M. we would less than ever want to
be parted, we see ourselves forced to dictate from writing-table
which measurements are in principle possible, that is, must be
possible in order to support adequately our reckoning system."
E. Schr\"odinger (\cite{Schrodinger35}, p.156) \end{quotation}}

`Observation' and `measurement results', which is the way by which
the physicists experience that which is {\it actual}, that which
is the case, are always necessary involved with the {\it
description} provided by the theory. The concept of object is a
creation, which works fairly well in our classical world. But, as
Einstein told to Heisenberg ``it is only the theory which can tell
you what can be observed." What is important to notice is that
only entities, which exist in the mode of being of actuality, can
be observed in classical physics. That `entities exist in the
world' is not a discovery of classical physics, but its basic
assumption.

{\smallroman
\begin{quotation}
``The classical tradition has been to consider the world to be an
association of {\small {\it observable objects}} (particles,
fluids, fields, etc.) moving about according to definite laws of
force, so that one could form a mental picture in space and time
of the whole scheme." P. Dirac (\cite{Dirac47}, preface to the
first edition, emphasis added)\end{quotation}}

Let us be clear about this point: one never observes {\it objects}
as such. An object is a conceptual machinery which is able to
unify our perceptions. It is a mental structure which is {\it
presupposed} in every experience which takes place in the domain
of classical thought. One does not encounter objects in the world,
one presupposes their existence and this allows us to create
experience, an experience which is for us, as physicists, an
expression of reality. Closer to our days we find in Bas van
Fraassen a strong defender of actuality.

{\smallroman
\begin{quotation}
``To be an empiricist is to withhold belief in anything that goes
beyond the actual, observable phenomena, and to recognize no
objective modality in nature. To develop an empiricist account of
science is to depict it as involving a search for truth only about
the empirical world, about what is actual and observable. Since
scientific activity is an enormously rich and complex cultural
phenomenon, this account of science must be accompanied by
auxiliary theories about scientific explanation, conceptual
commitment, modal language, and much else. But it must involve
throughout a resolute rejection of the demand for an explanation
of the regularities in the observable course of nature, by means
of truths concerning a {\small{\it reality beyond what is actual
and observable, as a demand which plays no role in the scientific
enterprise}}." B. van Fraassen (\cite{VanFraassen81}, pp.202-203,
emphasis added)\end{quotation}}

For us, there is no representable `actual' account of the world
voided of description. Actuality, when represented, is not left
without strong presuppositions. Representation takes place through
concepts and one must presuppose entities to even talk about such
actuality. A direct access to actuality presents us with
unavoidable paradoxes as a world of pure sensation remains outside
the limits of language and expression. Irineo Funes, as recalled
by Jorge Luis Borges, had thought of such a language but left it
aside for obvious reasons:

{\smallroman
\begin{quotation}
``Locke, en el siglo XVII, postul\'o (y reprob\'o) un idioma
imposible en el que cada cosa individual, cada piedra, cada
p\'ajaro y cada rama tuviera un nombre propio; Funes proyect\'o
alguna vez un idioma an\'alogo, pero lo desech\'o por parecerle
demasiado general, demasiado ambiguo. En efecto, Funes no s\'olo
recordaba cada hoja de cada \'arbol de cada monte, sino cada una
de las veces que la hab\'ia percibido o imaginado. Resolvi\'o
reducir cada una de sus jornadas pret\'eritas a unos setenta mil
recuerdos, que definir\'ia luego por cifras. Lo disuadieron dos
consideraciones: la conciencia de que la tarea era interminable,
la conciencia de que era in\'util. Pens\'o que en la hora de la
muerte no habr\'ia acabado de clasificar todos los recuerdos de la
ni\~{n}ez.

Los dos proyectos que he indicado (un vocabulario infinito para la
serie natural de los n\'umeros, un in\'util cat\'alogo mental de
todas las im\'agenes del recuerdo) son insensatos, pero revelan
cierta balbuciente grandeza. Nos dejan vislumbrar o inferir el
vertiginoso mundo de Funes. \'Este, no lo olvidaremos, era casi
incapaz de ideas generales, plat\'onicas. No s\'olo le costaba
comprender que el simbolo gen\'erico \small{{\it perro}} abarcaba tantos
individuos dispares de diversos tama\~{n}os y diversa forma; le
molestaba que el perro de las tres y catorce (visto de perfil)
tuviera el mismo nombre que el perro de las tres y cuarto (visto
de frente). [...]

Hab\'ia aprendido sin esfuerzo el ingl\'es, el franc\'es, el
portugu\'es, el lat\'in. Sospecho, sin embargo, que no era muy
capaz de pensar. Pensar es olvidar diferencias, es generalizar,
abstraer. En el abarrotado mundo de Funes no hab\'ia sino
detalles, casi inmediatos." J. L. Borges (\cite{Borges},
pp.489-490)\end{quotation}}

For us there are no `naked facts', a phenomenon comes from the
synthesis between the description ---which determines the conditions
of possibility to access a certain aspect of the Being--- and
experimental observation (the Being as exposed by the
description). These two elements interact with no preponderance of
one over the other, they are regarded by reality like two mirrors
with nothing in between.

For more than one century we have been looking at the theory
through an eyehole, we have seen only that which goes through the
door of actuality. We believe that potentiality, conceived as a
different mode of the being, is the key which might allow us to
enter the quantum domain.

\subsection{Quantum Mechanics as a Theory of the Potential}

The concept of potentiality has occupied a fundamental position in
the history of occidental thought. Its relation to actuality has
been one of the first, and maybe, still unresolved problems in
western philosophy. The common idea is that the real is reducible
to that which is `actual', from this position the conception of a
`potential non-actual' is denied. Aristotle criticized the
Megarians who stated that potentiality only exists in actuality,
his logic, however, was interpreted following these same steps.

Quantum mechanics was developed from a critical revision to the
idea of preexistence and it is in this very sense that it involves
an attempt to escape the limitations imposed by the classical
picture of the world in terms of actuality. This departure was
given by the mathematical language which Heisenberg developed as a
direct consequence of Planck's quantum postulate. The
philosophical guiding line was already developed by Mach as a
critical analysis of the metaphysical ideas of classical Newtonian
mechanics. Niels Bohr, contrary to Heisenberg and Pauli, wanted to
save, above all, the classical description, avoiding any type of
conceptual development. According to him: ``[...] the unambiguous
interpretation of any measurement must be essentially framed in
terms of the classical physical theories, and we may say that in
this sense the language of Newton and Maxwell will remain the
language of physics for all time." One might say that rather than
``suspended in language" we are ``stuck'' in (classical) language.

Heisenberg and Pauli, contrary to Bohr, seeked for new means of
expression. On the one hand, Wolfgang Pauli (\cite{Pauli94}, p.126) criticized the very
categorical pre-conceptions involved in the Kantian scheme:
``We agree with P. Bernays in no longer regarding the special
ideas, which Kant calls synthetic judgements a priori, generally
as the pre-conditions of human understanding, but merely as the
special pre-conditions of the exact science (and mathematics) of
his age."  His path was to study carefully the idea of space and
time in Alchemy and in Kepler's writings, as he stated in a letter
to Fierz on December 29, 1947: ``I find the time particularly interesting, when space and
time were \emph{not yet} up there and, indeed, the moment precisely \emph{before} this fateful operation. This is my reason for my study of Kepler."\footnote{Quoted from \cite{Laurikainen88}, p.202.} Pauli was seeking to develop the concept of reality, he certainly knew about the conceptual difficulty with which he was dealing and saw in the idea of complementarity a way to regain a picture of the world.\footnote{A possible development of this idea was investigated in \cite{deRondeCDI, deRondeCDII}.} Werner Heisenberg, on the other hand, developed the idea of \emph{potentia} as read from the {\it Timaeus} of Plato.

{\smallroman
\begin{quotation}
``In the experiments about atomic events we have to do with things
and facts, with phenomena that are just as real as any phenomena
in daily life. But atoms or the elementary particles {\small
\emph{are not as real}}; they form a world of potentialities or
possibilities rather than one of things or facts." W. Heisenberg
(\cite{Heisenberg58}, p.160, emphasis added)
\end{quotation}}

\noindent However, it is clear from his own writings that the idea
of potentiality was still thought in terms of actuality, as mere
possibility. This interpretation steams form the orthodox reading
of Aristotle which does not take into account the ontological
aspect of potentiality as a \emph{mode of the being}, independent
of actuality. Continuing the path laid down by Pauli and
Heisenberg the problem which we propose to resolve is the
following: \emph{how is it possible to think the real in terms of
potentiality?} It is clear for us however that the orthodox
interpretation of potentiality has clear limitations.

{\smallroman
\begin{quotation}
``Aristotle [...] created the important concept of {\small {\it
potential being}} and applied it to {\small {\it hyle}}. [...]
This is where an important differentiation in scientific thinking
came in. Aristotle's further statements on matter cannot really be
applied in physics, and it seems to me that much of the confusion
in Aristotle steams from the fact that being by far the less able
thinker, he was completely overwhelmed by Plato. He was not able
to fully carry out his intention to grasp the {\small
\emph{potential}}, and his endeavors became bogged down in early
stages." W. Pauli (\cite{PauliJung}, p.93)\end{quotation}}

We believe it was Pauli who had most clearly seen this path, as
noted in a letter to Carl Gustav Jung dated 27 February 1953:

{\smallroman
\begin{quotation}
``Science today has now, I believe, arrived at a stage where it can
proceed (albeit in a way as yet not at all clear) along the path
laid down by Aristotle. The complementarity characteristics of the
electron (and the atom) (wave and particle) are in fact
``potential being," but one of them is always ``actual nonbeing."
{\small {\it That is why one can say that science, being no longer
classical, is for the first time a genuine theory of becoming and
no longer Platonic.}}" W. Pauli (\cite{PauliJung}, p.93, emphasis
added)\end{quotation}}

\subsection{Classical Potentiality vs. Ontological Potentiality}

The idea of regarding actuality as the real is a heavy burden in
western thought which comes already from Aristotelian philosophy
and its cosmology. In quantum mechanics the notion of potentiality
was used for the first time by Heisenberg, his interpretation was
followed by several other approaches which maintained the same
idea of interpreting the quantum wave function as a {\it tendency}
or {\it propensity} to become actual.\footnote{Such are the
interpretations of Henry Margenau, Karl Popper, Constantin Piron,
Diederik Aerts and more recently by Mauricio Su\'arez.} All these
interpretations relied directly on the concept of actuality, their
problem has been to explain {\it how things become actual}, this
is why their definitions of potentiality or propensity find their
limit in the concept of actuality. In these interpretations actuality
remains that which is real and potentiality a secondary element by which one is able to
explain the actual. From the very start of our investigation we
have taken distance from such approaches by distinguishing
our own notion of potentiality which we have called: {\it
ontological potentiality} \cite{deRondeOP, deRondeCDII}. In
different opportunities we have stated that ontological
potentiality is a different `mode of the being' to that expressed
by actuality. According to us, the central point of this concept
is that it confronts us with the necessity of considering
potentiality as ontologically independent of actuality.

As a matter of fact, Aristotle himself had distinguished between two
types of potentiality. Firstly, he talked about a {\it generic
potentiality}: the potentiality of a seed that can transform into
a tree. It is important to notice that this idea of potentiality
presents actuality as its main goal, as a process which finds its
resolution in an actual state of affairs. `I have the potential
possibility of raising my hand' means that either {\it I will}
raise my hand or {\it I will not}. According to Agamben, this
generical sense is not that which interested Aristotle, who's
thought was concentrated in a different notion \cite{Agamben99}.
Aristotle was interested in discussing {\it potentiality as a mode
of existence}: the poet has the capacity of writing poems and of
not writing poems. It is not only the potentiality of doing this
or that thing but also the potentiality of not-doing, potentiality
of not being, of not passing to the actual. What is potential is
capable of being and not being. This is the problem of
potentiality: {\it the problem of possessing a faculty.} What do I
mean when I say ``I can", ``I cannot". Ontological potentiality is
a mode of existence which expresses \emph{power to do}, and
\emph{power not to do}, \emph{pure action} as well as \emph{pure
inactivity}.\footnote{In this sense we have taken distance from
Giorgio Agamben (\cite{Agamben99}, p.183) who provides a negative
interpretation to potentiality: ``[...] `To have a faculty' means
to have a privation. And potentiality is not a logical hypostasis
but the mode of existence of this privation." We wish to thank
Fernando Gallego for the many discussions regarding this important
point.}

The notions of tendency and propensity are thought in terms of a
process which has its final stage, its goal, in actuality. In this
sense, if they exist, it is only because of actuality, but they cannot
be thought without direct relation to the actual state of affairs.
It must be clear that ontological potentiality is not a tendency
nor a propensity, its definition does not rely in any way to what
{\it will be} the case in the future. Ontological potentiality
{\it is}, it exists in the present, here and now.

The concept of potentiality as a mode of existence has been used
implicitly or explicitly in the development of quantum mechanics.
As noted by Heisenberg: ``I believe that the language actually
used by physicists when they speak about atomic events produces in
their minds similar notions as the concept of `potentia'. So
physicists have gradually become accustomed to considering the
electronic orbits, etc., not as reality but rather as a kind of
`potentia'." (p.156) Maybe the most interesting example of an
implicit use of these ideas has been provided by Richard Feynmann
in his path integral approach. Even though Feynman talks about
calculating probabilities, he thinks in terms of existent
potentialities. Why, if not, should we take into account the
mutually incompatible paths of the electron in the double slit
experiment? His approach takes into account every path as existent
in the mode of being of potentiality, there where the constrains
of actuality cannot be applied (see \cite{FeynmanHibbs65} section 1.3). We will return to this point later
in section 6.1.

\section{The Quantum Wave Function in Terms of `Interacting
Faculties'}

Concepts are creations, they are not God given. And just like the concept of `entity' was created, it is in principle possible to think in a different concept which could describe physical reality. Our investigation has analyzed exactly this problem: how can we develop a concept which brings into stage that of which quantum mechanics is talking about in terms of an objective account of physical reality? In order to solve this problem we have introduced in \cite{deRondeCDI, deRondeCDII} the concept of `faculty'.

An experimental arrangement is nothing but the condition of
possibility for an action to take place, it creates the {\it power
to} perform an experiment. In quantum mechanics we are faced with
the choice of mutually incompatible experimental arrangements, each of
which expresses a given capability, this `power to do' is,
according to us, something which exists in the world, it is this
`ontological element' what we call a \textbf{faculty}. The
principles of indetermination, complementarity and superposition
determine the notion and meaning of `faculties', just like the
principles of existence, non-contradiction and identity provide
the constraints for a proper determination of the concept of
`entity'. Our aim in this section is to go deeper into the quantum
principles and explain more clearly, if possible, what do we mean
with the concept of `faculty'.

\subsection{The Mode of Being of Faculties: Indetermination and Complementarity}

In order to understand what we mean by a faculty we need to have
in mind two general rules which are not so easy to follow.
Firstly, we have to forget about a direct reference to entities,
and even though language forces us into this Socratic trap we
should avoid from now on committing ourselves to this particular
view. Secondly, in order to avoid thinking in the old terms of
potentiality, in terms of tendency, in terms of possibility, we
should always think of faculties as existents in the present
tense, as an element of reality which exists here and now, independently of
what will actually be the case. With these two ideas in mind we
are now ready to continue.

Faculties are a machinery which can allow us to compress the
quantum experience into a picture of the world, just like entities
such as particles, waves and fields, allow us to do so in classical
physics. The mode of being of a faculty is potentiality, not
thought in terms of possibility (which relies on actuality) but
rather in terms of ontological potentiality, as a mode of
existence. I have the faculty of {\it raising my hand}, which does
not mean that `{\it I will} raise my hand' or that `{\it I will
not} raise my hand'; what it means is that here and now I possess
a faculty which exists in the mode of being of potentiality,
independent of what will happen in actuality. Faculties do not
exist in the mode of being of actuality, faculties are not actual
existents, we cannot ``see" faculties, we can only {\it experience} with
them. It is important to notice there is no difference in this
point with the case of entities, as we have discussed earlier: we
cannot ``see" entities either. Entities, in classical mechanics,
as well as faculties, in quantum mechanics, are the {\it basic
presuppositions} needed for the determination of the classical and
quantum experience, they act as the machinery which is able to bring together observation and measurement.

Faculties should not be regarded as equivalent to a process, there is
no need of a lapse of time for a faculty to exist. Faculties exist
instantaneously, in the mode of being of ontological potentiality. The {\it process} is that through which we access the faculties, in the same way we access entities through an examination of their properties. Entities exist {\it per se}, as {\it essences}, independently of the rest of the world, they are non-contextual existents. Faculties, on the other hand, are explicitly determined in relation to what we can do in a definite {\it state of affairs}; i.e. they are relational {\it contextual existents}.  The notion of {\it complementarity} plays a central role here, understood in this case, not as bringing together different incompatible representations, but rather, as providing the constrains under which faculties exist.

A faculty maintains a logic of actions and relations which do not necessarily take place in actuality, a faculty \emph{is} and \emph{is not}, \emph{here} and \emph{now}. Heisenberg's principle must be understood in this case as providing the mathematical expression of this basic character of faculties which refers to its being {\it indetermined}. The difference with entities in classical physics regards the way in which this experience is produced, the conditions which allow us to experience with faculties are certainly different.

Faculties are indetermined and contextual existents. A faculty is structured always by a certain `power to do', a power which relates to the configuration of relations in a given {\it state of affairs}. I possess the faculties of {\it running} and {\it swimming}, but in order for these faculties to exists, I must be either in a place where I can run, or in a place where I can swim. I can say: ``\emph{I can} swim (here and now)" only if, given the state of affairs, I am able to do so. In order to swim I obviously need to be in a place where I can swim, like for example in a swimmingpool. This has nothing to do with the fact that in the near future I choose either to swim or not to swim while I'm in the swimmingpool. In a swimmingpool however, I am not able to run, just in the same way that I am not able to swim in the street. In our earlier terms the context determines the existence of the faculty explicitly.

\subsection{Understanding the Notion of `Superposition' in terms of a Faculty}

A basic question which we have posed to ourselves at the beginning of our trip regards the meaning of a superposition. What does it mean to have a superposition $|\psi_{B}\rangle = \alpha |up\rangle + \beta |down\rangle$? How can we most clearly expose it conceptually and relate it to physical reality? Our theory of faculties has been developed in order to answer this particular question and is an appendix of our earlier distinction between perspective and context.

The entanglement between the idea of entity and the structure of the quantum formalism was discussed above, placing the choice of the basis as an active constituent of that which is discussed in a definite context, namely, a superposition. It is important to notice that in relation to the active status of the basis in the superposition, given the `x basis'  we obtain a faculty, call it $F_{x}$, $|\alpha_{x}\rangle + |\beta_{x}\rangle$, while a rotation to the `y basis' gives place to a different faculty $F_{y}$, $|\alpha_{y}\rangle + |\beta_{y}\rangle$. These two mathematical expressions $F_{x} = \frac{1}{\sqrt{2}}[|\alpha_{x}\rangle + |\beta_{x}\rangle]$ and $F_{y} = \frac{1}{\sqrt{2}}[|\alpha_{y}\rangle + |\beta_{y}\rangle]$ give place to different {\it incompatible} existents. In this sense, {\it incompatibility} is a central feature of faculties.

We understand a superposition as encoding the state of the faculty and its power. The notion of state of a faculty goes against the basic principles of Aristotelian logic. Firstly, it does not exist in actuality, we cannot see a faculty in the same way we see an object. We understand the state of a faculty as existing in the realm of ontological potentiality. Secondly, the elements of that which constitutes the state of a faculty violates the principle of non-contradiction. The logical structure of a faculty ---given mathematically by the superposition--- is such that a property and its negative exist at one and the same time, just in the same way, when I have the faculty of raising my hand both actions (`raising my hand' and `not raising my hand') co-exist in the definition itself of having a faculty. The faculty is {\it sustained activity}, something which {\it is} and {\it is not}. Finally, if thought in terms of faculties the notion of {\it identity} simply losses its meaning. A faculty is not a substantive of which one can predicate certain properties. A faculty is a sustained verb, sustained activity, and it makes no sense to talk of verbs as having identity or individuality. Would it make sense to ask if the faculty of swimming can be one and the same through time? This is simply a badly posed question. One can make this question with respect to entities because entities exist as essences, and in this sense there is something which remains the same and equal to itself. In the case of faculties we do not deal with essences, but with {\it pure relations}.

\subsection{Faculties Instead of Entities}

Our strong thesis is that we have been stating the wrong questions
to quantum mechanics, we have been always asking about `entities'
while quantum mechanics can only answer questions which have to do
with `faculties'. An entity, as thought in physical terms, is
governed by the principles of classical (Aristotelian) logic:
principle of existence, principle of identity and principle of
non-contradiction. A faculty can be thought in terms of the principles which
give place to quantum theory, namely: Heisenberg's principle of
indetermination, Bohr's principle of complementarity and the
superposition principle. In the ontology we are discussing there
is only {\it active relations}, a logic of action in contraposition to
the statical logic of Aristotle. How to think in terms of this
logic is not obvious and might be regarded as the most difficult
task of our time. We have to learn to think in terms of change and process. Our
strong claim is that, {\it just like entities exist in the realm
of classical physics, faculties exist in realm of quantum
physics.} Instead of a logic of essences which refers to entities,
we have a logic of actions which refers to faculties.\footnote{In this line of thought we call the attention to the work of Bob Coecke and Sonja Smets regarding the dynamical development of operational quantum logic, see \cite{CoeckeSmets04} and \cite{SmetsPHD}.}

In terms of our distinctions between perspectives and contexts, a superposition, $|\psi_{B}\rangle$, can be thought as being ``the state of a faculty''. The perspective, $\Psi$, can be thought as pure {\it potentia}, in the sense of pure relational activity, as describing pure, non-actualized relations between faculties. It should be noted that the term {\it potentia} to which we refer should be understood not in terms of Aristotle but rather in relation to Spinoza, as a {\it power to do}, {\it power to affect}. In this framework there are no entities whatsoever, entities appear only in later stages, when we destroy through our choices, the basic characters of the quantum description and we
impose the classical structure. After this fateful operation is
produced {\it indetermination} is translated into {\it uncertainty}, the
ontological {\it incompatibility} of properties into a discursive
{\it complementarity} of classical representations, and finally,
{\it superpositions} are simply forgotten and read as a mathematical
weirdness which gives place to algorithmic results.

In order to put everything which have been exposed until now we
present the following diagram:\\

\begin{tabular}{|c|c|c|c|c|}

\hline
& \tiny{\textbf{PERSPECTIVE}} & \textbf{\tiny{HOLISTIC}} & \tiny{\textbf{REDUCTIONISTIC}} & \tiny{\textbf{MEASUREMENT}} \\
&  & \textbf{\tiny{CONTEXT}} & \tiny{\textbf{CONTEXT}} & \tiny{\textbf{RESULT}} \\
\hline
\tiny{\textsl{MODE OF BEING}} & \footnotesize{{\it potential}} & \footnotesize{{\it potential}}  & \footnotesize{{\it possible/probable}} & \footnotesize{{\it actual}} \\
\hline
\tiny{\textsl{MATHEMATICAL EXPRESSION}} & $\Psi$ & $|\psi_{B}\rangle$ & $|\psi_{B}\rangle$ & $\alpha_{k}$, $|\alpha_{k}\rangle$ \\
\hline
\tiny{\textsl{CONCEPTUAL EXPRESSION}} & \footnotesize{{\it active relations}} & \footnotesize{{\it superposition}} & \footnotesize{{\it ensemble}} & \footnotesize{{\it single term}}\\
\hline
\tiny{\textsl{PROPERTY}} & -- & \footnotesize{{\it holisic/non-Boolean}} & \footnotesize{{\it reductionistic/Boolean}} & \footnotesize{{\it actual}} \\
\hline
\tiny{\textsl{DESCRIPTION IN TERMS OF}} & \footnotesize{{\it potentia}} & \footnotesize{{\it faculty}} & \footnotesize{{\it possible entities}} & \footnotesize{{\it actual entity}} \\
\hline
& \footnotesize{{\it indetermination}} & \footnotesize{{\it indetermination}} & \footnotesize{{\it existence}} & \footnotesize{{\it existence}} \\
\tiny{\textsl{LOGICAL PRINCIPLES}} & \footnotesize{{\it complementarity}} & \footnotesize{{\it complementarity}} & \footnotesize{{\it non-contradiction}} & \footnotesize{{\it non-contradiction}} \\
& \footnotesize{{\it superposition}} & \footnotesize{{\it superposition}} & \footnotesize{{\it identity}} & \footnotesize{{\it identity}} \\
\hline
\end{tabular}\\

\subsection{...What can be Observed?}

The problem of measurement in quantum theory has been a great matter of debate. As noted already, measurements become a completely subjective structure when related to entities, in quantum mechanics entities exist only because we measure.\footnote{This can be justified in many ways, maybe the most clear for us remains the constrains imposed by KS theorem, which makes clear the fact that one has to choose different sets of mutually incompatible properties, this choice determines explicitly the entity under study.} Our investigation has been guided by the words of Einstein who's echoes have reached our days: ``It is only the theory which can tell you what can be observed''. As we noted already, our intention is to recover an objective account of the states of affairs discussed by quantum theory. In this sense, one of the most important tasks which we have assumed is the analysis of the meaning of measurement in quantum mechanics. In particular, we must be able to give a proper account of the measurement process in relation to faculties, i.e. we must provide the conditions under which it is meaningful to talk about a ``measurement of a faculty''.

In classical mechanics we observe entities which exist as elements of an essentially static structure\footnote{In classical physics the static structure regards the logical scheme already put forward by Aristotle. As noted by Verelst and Coecke: ``[...] change and motion are intrinsically not provided for in [the Aristotelian] framework; therefore the ontology underlying the logical system of knowledge is essentially static, and requires the introduction of a First Mover with a proper ontological status beyond the phenomena for whose change and motion he must account.'' Quoted from \cite{VerelstCoecke}, p. 172.} and observation appears as bringing into stage that which exists in actuality. In quantum mechanics things take place in a very different way, we observe faculties which make themselves present through action and change. {\it In quantum theory we only measure shifts of energy, change, processes}. This fundamental point was already noticed by Nancy Cartwright:

{\smallroman
\begin{quotation}
``It makes good sense to take energy transitions as basic for the interpretation of quantum mechanics. For it is only through interchanging energy that quantum systems interact and can register their interactions in a macroscopically observable way. In a very well-known argument against reduction of the wave packet, Hans Margenau has urged that all measurements are ultimately measurements of position. But this should be pushed one step further. All position measurements are ultimately measurements of energy transitions. No matter that a particle passes by a detector ---the detector will not register unless it exchanges some energy with the particle. The exchange of energy is the basic event that happens in quantum mechanics; and the basic event whose effects are theoretically described and predicated.'' N. Cartwright (\cite{Nancy}, p.55)\end{quotation}}

The concept of faculty confronts a problem which does not find an answer in terms of entities. How can we think of something which is {\it different} every time it is realized in an experimental procedure but rests simultaneously one? Let's imagine that we are in the shore of a river, its full of stones. We grab a stone and through it into the river, we grab another stone and through it, and then another one. Each one of the stones is different, we through them from different places at different times and even the lake changes as we add stones to it which were not there before. The question we should ask is what can be generalized in this process? Every action involves a singularity because there are different stones involved, the sun crosses the sky, I get older as time passes by. The abstraction we can do in order to generalize that which we have described deals with the process itself, that to which we can refer as being the same is `the action of throwing rocks'. In order to find regularities we need to shift from the subject to the verb. That which is the same but has no reference to something is a faculty. A faculty is observed through a process. The stones are not the same, neither the lake, that which remains ``the same'' is the action itself. A repetition of a difference.

The orthodox interpretation presents the superposition as referring to the electron itself, to the probability of finding a particular property of this entity. Thus, within this interpretation, the superposition encodes the properties of a system. Our interpretation in terms of faculties presents the superposition as referring to a certain faculty, which {\it I have} in relation to the experimental arrangement, a `power to do' which is encoded in a mathematical expression. The faculty is observed through a process, a shift in energy. Observation takes place through the shift of energy within a given state of affairs. Objectivity is regained in quantum measurements when we forget about entities and discuss in terms of faculties. Just like entities exist even when there is no light to see them, faculties exist in the world regardless of observation and measurement outcomes. Just in the same way that entities appear to us through contemplation, and remain in the dark when light does not shine upon them, faculties can be observed through the shifts of energy and remain unknown when change is forgotten. In quantum mechanics only change, shifts of energy, are taken into account, the quantum postulate does not only imply a different way of acquiring sense data, it is the basic cornerstone of a definition of a new experience.

It is important to remark however that these examples are only limited in their use, more specifically it should be noticed that the capability of observing actions within the classical scheme is continuous, and not discrete as in the case of quantum mechanics. In classical mechanics we observe continuous process. The notion of continuity is here of major importance and has not been investigated adequately.\\

\begin{tabular}{|c|c|c|}

\hline
& \textbf{Classical Mechanics} & \textbf{Quantum Mechanics} \\
\hline
\textsl{OBSERVATION} & \footnotesize{{\it Continuous path.}} & \footnotesize{{\it Discrete shift.}}\\
\hline
\textsl{MEASUREMENT} & \footnotesize{{\it Properties}} & \footnotesize{{\it Process (energy shift)}}\\
\hline
\textsl{Objective account of physical} & \footnotesize{{\it Entities}} & \footnotesize{{\it Faculties}}\\
\textsl{reality in terms of ...} &  & \\
\hline
\end{tabular}
\\

\section{Interpreting Quantum Paradoxes}

Our theory of faculties will be interesting, only in the case it
is able to provide the formalism of quantum mechanics with a
picture (an anshaulich content) which describes the experience
provided by quantum theory in an elegant way. A way which provides
a deeper understanding of what is going on according to quantum
theory. Off course, our idea is not to provide understanding in
classical terms, like for example Bohm's causal interpretation or
many worlds intend to do. There are many hidden presumptions and
intentions which can make us differ in our choice for a definite
interpretation. Justification of the choice we make might differ,
and even though empirical success remains basic, also beauty,
simplicity and the conceptual richness to provide new questions
(rather than answers) are always implicitly or explicitly taken
into account by physicists. We hope that our conceptual scheme is
able to shed new light regarding the question: what is quantum
mechanics talking about? In this paper we will focus in
interpreting, through our theory of faculties, two of the most discussed
experiments of quantum mechanics, namely, the double slit experiment
and Sch\"odinger's cat.

\subsection{The Double-Slit Experiment: Complementary Representations}

The double-slit (DS) experiment was one of the first to expose the
paradoxical character of the quantum formalism with respect to
classical physics. Niels Bohr and Albert Einstein discussed in
many occasions the possible interpretation of this thought experiment,
which shows that the `same' quantum wave function provides
information of incompatible classical representations, such as
those of `particles' and `waves' (\cite{WZ}, p.9). There is a weird entanglement
between the entities involved (particles and waves) and the mathematical formulation
which represents them. The most important assumption involved in the DS
experiment is that the quantum wave function makes reference to
some kind of entity. It is this hypothesis, which remains untouched at the basis of our
classical reasoning about physical reality, which has not been adequately discussed. There are further
presuppositions involved however in this experiment which we would
like to analyze, this we'd like to do in terms of an {\it ad
absurdum} proof, whose
hypothesis are the following:\footnote{This is in analogous way to Diederik Aerts' reading of EPR \cite{Aerts85}.}\\

{\smallroman
\begin{quotation}
\noindent\emph{$H_{1}$ (entity existence)}: There is some kind of
entity
which we are studying through the DS experiment.\\

\noindent\emph{$H_{2}$ (quantum representation)}: The quantum wave
function $\Psi$ respects the rules provided by quantum mechanics
and represents a feature of physical reality as
exposed by the experiment.\\

\noindent\emph{$H_{3}$ (empirical consistency)}: Observation
discovers some
unknown but preexistent property of the entity under study.\\

\noindent\emph{$H_{4}$ (objective consistency)}: The entity as
represented by our physical theory
exists independently of observation.\\
\end{quotation}}

The DS experiment shows most clearly that if one assumes all these
hypotheses simultaneously one can logically deliver a
contradiction, namely, that the electron, as represented by the
quantum wave function $\Psi$, `is a particle' ($|\psi_{part}\rangle$) and `is a wave' ($|\psi_{wave}\rangle$). In our terms, the
idea behind the curtains is that the perspective $\Psi$ is
something of which the {\it reductionistic contexts},
$|\psi_{part}\rangle$ and $|\psi_{wave}\rangle$, are mere
representations, detached observations. The quantum wave function represents the
`electron' which is presupposed to exists in physical reality, but
its existence is contradictory. Depending of the experimental
set-up the electron behaves {\it as if} it was a particle, or {\it
as if} it was a wave. Thus, it is not possible to presuppose the
existence of the quantum wave function $\Psi$ as an entity with
definite (non-contradictory) properties. There is no analogue in
classical thought of this experience.\footnote{It is important to
notice that in classical experience there are also incompatible
experimental arrangements. Diederik Aerts proposed to discuss in relation to this, a piece of wood
which has the property of being `burnable' and the property of
`floating' \cite{Aerts81}. The experiments which express the property are
mutually incompatible, however the important point is that in the
classical scheme one can think of both properties as existents of
the entity under study without getting into contradictions.
Experiment appears here only as discovering the properties of the
entity not as constituting them.} The experiment makes clear that
there is at least one of these hypothesis which must be left
aside.

Most discussions have been centered in leaving aside hypothesis
$H_{2}$, $H_{3}$ and $H_{4}$. For example, Bohmian mechanics tries
to change $H_{2}$ and proposes instead a new theory, with rules
closer to classical ideas. Also GRW changes the Schr\"odinger
equation of motion to a non-linear one in order to explain clearly
the relation between the mathematical scheme and experience
without quantum jumps. For reasons which might be already clear
form earlier sections, we are mostly interested in Bohr's proposal
which attacked mainly $H_{3}$ and $H_{4}$. Bohr's idea is to take
{\it quantum mechanics as a regulative theory of classical
representations} and that, by changing the notion of objectivity ---now considered in terms of intersubjective agreemnet---
one might retain the classical description needed to describe
(classical) phenomena. As discussed above, Niels Bohr created the
concept of complementarity in order to bring together incompatible
(classical) representations. Complementarity, according to Bohr,
is a {\it regulative principle} which allows to secure objectivity
and our classical discourse about phenomena. Heisenberg's
principle is interpreted in this context as {\it uncertainty
relations}, as providing the constraints of knowledge of the
properties under study. This gnoseological interpretation presents
quantum mechanics as providing a secure ground to {\it talk about
Nature}, according to classical physics, avoiding at the same
time the unconfortable discussion regarding the relation between physical reality and quantum mechanics.

$H_{1}$ remains such a deep presumption of classical physical
ideas that most of the discussions that one encounters do not even
mention $H_{1}$ as hypothesis. The basic assumption that the
quantum wave function describes some kind of entity remains
untouched in every discussion. But, is it necessary to make such
presuppositions in quantum physics when we know that quantum
mechanics was born from the very departure of the classical
description of Nature in terms of entities? As discussed above our
main idea is to leave aside $H_{1}$ and show that one can still
make sense of the double slit experiment. We have to stop thinking
in terms of `entities' which exist in the world, in its place we
propose to discuss this experiment of quantum mechanics
in terms of `faculties'. Let's see how this works out.

It is important to remark that in the DS experiment the state of affairs is changed drastically by adding an apparatus; i.e. the plate which covers one of the slits. The state of affairs which refers to the two slits open determines
the faculty of `producing an interference pattern'. Just in same
way that if I am in a swimmingpool I have the faculty of swimming
and I can then say: ``I can swim (here and now)", when the two
slits are open, the experimenter can say: ``The shift in energy of this state of affairs produces an `interference pattern' in the photographic plate (here and now)", as
a secondary inference, in this case we are allowed to talk about
waves. On the other hand, if we close one of the slits the
experimenter can refer to ``The shift in energy of this state of affairs produces a `Gaussian
pattern' in the photographic plate (here and now)", and in this case we are
allowed to speak about particles. The new state of affairs, the
new context, determines a new faculty (which was hidden in the
perspective). In this case the experimenter can say: ``I can
produce a Gaussian pattern in the photographic plate (here and now)" just
like I can say: ``I can run" if I go out of the swimmingpool and
jump into the street.

Quantum physics does not presuppose the existence of particles nor
waves, these are just derivative concepts which are produced by
the given state of affairs and several interpretational cuts which
have been discussed in \cite{deRondeCDII}. Quantum mechanics does
not talk about particles nor waves (swimmers nor runners), it
talks about faculties. Off course when I am in a certain state of
affairs I can only perform those experiments which are then
brought into existence. A particular state of affairs may always
turn {\it incompatible} a different state of affairs. In the same way,
running and swimming are incompatible faculties, each faculty
presupposes a state of affairs which precludes the possibility of
existence of the other. If I have the faculty of swimming, I must
be obviously in some place with water, but in such place I cannot
run. Contrary to this if I am able to run, I should be for example
in the street, but then, obviously I will not be able to swim.

We believe that the language which we are putting forward is more
appropriate to discuss quantum experiments, it can even open the
doors of new experiments which have been hidden by the classical
(entity dependent) description. The weirdness of the double slit
experiment appears when one wants to stop talking about faculties
and starts talking about entities, instead of talking about the
faculty of running or the faculty of swimming one wishes to return
to the classical realm and talk about swimmers and runners,
particles and waves.

\subsection{Schr\"odinger's Cat: Superposition of Properties}

Schr\"odinger's cat experiment is one of the best examples of what
happens when one mixes the quantum formalism with the classical
language \cite{Schrodinger35}. The paradox appears when we force the results of the quantum
formalism into the classical language. The cat cannot be dead or
alive simply because a cat is an entity, and as such, it
presupposes the classical description. In the classical
description every property is determined and cannot relate to
others in the way the quantum formalism indicates.

Now, how do we interpret this experiment through our theory of
faculties? Here, we must forget about cats and electrons
possessing properties in the actual mode of existence, this is the
trap into which the classical language forces us inn. What we have is
the faculty, call it $F_{S_{x}}$, of `having spin in the x-direction' given that we have
our Stern-Gerlach apparatus aligned in the x-direction. We can
only refer to the faculties which arise from this given state of
affairs.

In quantum mechanics we get into weird contradictions
when we choose to talk about entities. In quantum mechanics there
is no preexistent property such as spin, simply because there is
no entity which has this property. In quantum mechanics a property is an answer to a question which relates to a shift of energy in a given state of affairs. One gets into contradictions if
one states that an electron, without proper reference to the
experimental arrangement (state of affairs), has the property of
having spin in the x-direction.

The incompatibility is however of a different degree to that of
the double slit experiment. In this case the incompatibility is
not between classical representations which correspond to the
level of reductionistic contexts but rather between superpositions
or improper mixtures which pertain to the holistic context. The
distance lyes in the fact that a superposition, contrary to a
particle or a wave, cannot be subsumed into the presuppositions of
the classical description. This is an incompatibility not between
classical representations but between purely quantum
representations.

\subsection{Classical and Quantum Experience}

In our terms, the incompatibility present in the DS
experiment is between what we have called `reductionistic
contexts'. Even though these reductionistic contexts can not be
thought as providing different views or representation of one and
the same entity, each of them, by themselves can be though as
exiting in terms of entities, i.e. in terms of particles or in
terms of waves (proper mixtures). There is an intrinsic difference
between the DS experiment and Schrodinger's cat experiments. This
latter experiment goes a step further and discusses the incompatibility of
holistic contexts, of superpositions, which
can never be thought in terms of entities, even when taken into
account separately. It is this difference which draws the subtle line between quantum and classical experience.

Richard Feynman \cite{Feynman65} referred to the double slit
experiment in the following terms: ``We choose to examine a
phenomenon [the double slit experiment] which is impossible,
absolutely impossible, to explain in a classical way, and which
has in it the heart of quantum mechanics. In reality, it contains
the only mystery." We do not agree with this statement, as we saw above this is only the tip of the
iceberg, there are much deeper problems into which quantum
mechanics confronts us.

As a final resume of the analysis we have been proposing:\\

\begin{tabular}{|c|c|c|}

\hline
& \tiny{\textbf{DOUBLE-SLIT EXPERIMENT}} & \tiny{\textbf{SCHR\"ODINGER'S CAT EXPERIMENT}} \\
\hline
\tiny{\textsl{ANALYSIS}} & \footnotesize{{\it Gnoseological}} & \footnotesize{{\it Ontological}} \\
\hline
& \footnotesize{{\it Complementary relations:} between classical} & \footnotesize{{\it Principle of Complementarity}: relates q-properties.}\\
\tiny{\textsl{PRINCIPLES}} &  \footnotesize{representations (wave/particle).} & \footnotesize{{\it Principle of Indetermination:} mode of}\\
\tiny{\textsl{AND}} & \footnotesize{{\it Uncertainty relations:} secures the limits of} & \footnotesize{existence of q-properties.} \\
\tiny{\textsl{RELATIONS}} & \footnotesize{complementary relations.} &  \footnotesize{{\it Principle of Superposition:} composition}\\
& \footnotesize{ The {\it superposition} is thought it terms of a {\it wave}.} & \footnotesize{of q-properties.}\\
\hline
\tiny{\textsl{SCHEMES}} & \footnotesize{{\it Perspective} $\Psi$ relative to the} & \footnotesize{Single {\it holistic context}}\\
& \footnotesize{ {\it reductionistic contexts} $\psi_{part}$ and $\psi_{wave}$.} & \footnotesize{$\psi_{B}=|up\rangle + |down\rangle$.}\\
\hline
\tiny{\textsl{CLASSICAL}} & \footnotesize{Talk {\it as if} it was {\it particle} or {\it wave};} & \footnotesize{{\it Cannot} talk in terms of {\it entities}.} \\
\tiny{\textsl{DESCRIPTION}} & \footnotesize{i.e. in terms of {\it entities}.} & \\
\hline
\tiny{\textsl{QUANTUM}} & \footnotesize{Changing the state of affairs by {\it adding}} & \footnotesize{The same state of affairs}  \\
\tiny{\textsl{DESCRIPTION}} & \footnotesize{ {\it an apparatus} produces different {\it faculties}.} &  \footnotesize{does not change the {\it faculties}.} \\
\hline
\end{tabular}

\section{Conclusions}

Quantum mechanics and classical mechanics do not talk about
different worlds, but rather, provide complementary descriptions
of one and the same world, they express reality through different
descriptions which have in their heart incommensurable relations.
Classical mechanics might be regarded as that which refers to
actuality through the principles of classical Aristotelian logic,
quantum mechanics, on the other hand, can be seen as providing a
description of the potential as a mode of the being, through the
principles of indetermination, superposition and complementarity
\cite{deRondeCDI, deRondeCDII, deRondeUQMCD}.

According to our view, quantum mechanics talks about faculties and their relations to the world. $\Psi$ is an expression of the
condition of possibility to perform a certain experiment. The
faculty describes a level which does not pertain to \emph{things}
but rather to \emph{potential actions}. A faculty is expressed in terms of an
\emph{objective state of affairs} through the shifts in energy observed in measurement interactions. Faculties
\emph{exist}; i.e. they are in the world just like ``things" are.
Classical physics is the study of the world as constituted by
entities which exist in the mode of being of actuality. Quantum
physics is the study of the world as constituted by faculties
which exist in the mode of being of potentiality. This should be
understood as a step forward in the level of abstraction regarding
our understanding of reality.

The price to pay, if we are willing to recover the objective character of quantum mechanics is to leave aside the concept of entity and with it, the whole classical description of the world. The most important test regarding this approach remains, as with any physical theory, the possibility to determine a new experience. It will be important not only to understand well known experiments, but also to find out about new ones, experiments which have not been thought until now, and which our theory of faculties can help to develop.

\section*{Acknowledgements}

I wish to thank Graciela Domenech and Federico Holik for a careful reading of earlier drafts and discussions on the many subjects of this paper. This work was partially supported by Projects of the Fund for Scientific Research Flanders G.0362.03 and G.0452.04.

\end{document}